\documentclass[preprint]{jpsj2}
\usepackage{amsmath}
\usepackage{bm}

\def\runauthor{Takeshi Yoshida and Mitsusada M. Sano}

\title{%
Numerical Simulation of Vortex Crystals and Merging\\
 in $N$-Point Vortex Systems with Circular Boundary}

\author{%
Takeshi YOSHIDA\thanks{E-mail address: tyoshida@phys.h.kyoto-u.ac.jp}
and Mitsusada M. SANO
}

\inst{%
Graduate School of Human and Environmental Studies,
 Kyoto University, Kyoto 606-8501
}

\recdate{\today}

\abst{%
In two-dimensional (2D) inviscid incompressible flow, low
 background vorticity distribution accelerates intense vortices (clumps)
 to merge each other and to array in
 the symmetric pattern which is called ``vortex crystals'';
they are observed in the experiments on pure electron plasma and
 the simulations of Euler fluid.
Vortex merger is thought to be a result of negative ``temperature''
 introduced by L. Onsager. Slight difference in the initial distribution
 from this leads to ``vortex crystals''.
We study these phenomena by examining $N$-point vortex systems
 governed by the Hamilton equations of motion.
First, we study a three-point vortex system without background distribution.
It is known that a $N$-point vortex system with boundary
 exhibits chaotic behavior for $N\geq 3$.
In order to investigate the properties of the phase space structure of 
 this three-point vortex system
 with circular boundary, we examine the Poincar\'e plot of this system.
Then we show that topology of the Poincar\'e plot of this system
 drastically changes when the parameters,
 which are concerned with the sign of ``temperature'', are varied.
Next, we introduce a formula for energy spectrum of
 a $N$-point vortex system with circular boundary.
Further, carrying out numerical computation, we reproduce a vortex
 crystal and a vortex merger in a few hundred point vortices
 system. We confirm that the energy of vortices is transferred
 from the clumps to the background in the course of vortex crystallization.
In the vortex merging process, we numerically calculate the energy
 spectrum introduced above and confirm that it behaves as
 $k^{-\alpha},(\alpha\approx 2.2-2.8)$
 at the region $10^0<k<10^1$ after the merging.
}

\kword{%
vortex, Poincar\'{e} section, energy spectrum, vortex dynamics,
 vortex crystals, vortex merger, numerical simulation, negative temperature
}

\begin{document}
\maketitle

\clearpage

\section{Introduction}
It is well known that strongly magnetized pure electron plasma is
approximately equivalent to 2D inviscid incompressible fluid. 
In a strong uniform magnetic field $\bm{B}_0=B_0\hat{\bm{z}}$
 the drift velocity of the guiding center of plasma is given by
$\bm{v}(x,y)=-\nabla \phi(x,y)\times{\hat{\bm{z}}}/B_0$,
where $\phi$ is the electrostatic potential and  
$\hat{\bm{z}}$ is a unit vector normal to the plane of the flow.
Thus, this flow of the drift velocity
 is incompressible, i.e., $\nabla \cdot \bm{v}=0$.
 The vorticity of this velocity field is given by
$\bm{\zeta}(x,y)=\nabla\times\bm{v}=\hat{\bm{z}}{\nabla}^2\phi/B_0$.
Using the Poisson equation, the vorticity of this velocity field is shown
 to be proportional to the electron density.
 Therefore, the vorticity is governed by the following equations,
\begin{align}
\frac{\partial \zeta}{\partial t}+\bm{v}\cdot\nabla \zeta&=0,
&\bm{v}&=\hat{\bm{z}}\times\nabla\Psi,
&\zeta &= {\nabla}^2\Psi. 
\end{align}
Here $\zeta(x,y)=(e/\varepsilon_0 B_0)n(x,y)$
 ( $n(x,y)$ is the electron density )
 and $\Psi(x,y)=(\varepsilon_0/e)\phi(x,y)$
 is the stream function. $\varepsilon_0$
 is the dielectric constant in vacuum.
Thanks to the above relation, the behavior of vortices in
 incompressible Euler fluid is
experimentally realized on non-neutral plasma. 
The vortex merging process and the motions of point vortices 
 in vacuum or in the background vorticity distribution
 were experimentally observed.~\cite{2,3,4}

Theoretically,
 stability of point vortices array had been studied.~\cite{Saffman,Kurakin}
Furthermore equilibrium states of $N$-point vortex system had been also
 investigated.~\cite{kida,Ralph,Chavanis,Ons}
The equilibrium states of this $N$-point vortex system
 under various initial conditions were obtained
 by using statistical mechanical method.
The theory using statistical mechanics predicted
 that the vortices of this system merge one after another
 to be a large vortex in equilibrium.
This phenomenon occurs when the ``temperature'' of point vortices is negative.
The positive ``temperature'' leads to merger between vortices of opposite sign,
 or with boundary.~\cite{Ons}

However, about ten years ago, a new phenomenon was observed
 in the experiment by Fine \textit{et al}.~\cite{Fine}
Under certain initial conditions,
 vortex merging is ``cooled'' and a few intense vortices
 (which is called ``clump'' in this paper to distinguish a point vortex)
 form a regular lattice in a low background vorticity distribution. 
This phenomenon which is called ``vortex crystals'' is the highlight
 of this field.
This vortex crystals are investigated from various aspects.
For example, a power law decay of the number of intense vortices of this
 system, namely $N\sim t^{-\xi}, \xi=0.2-0.7$
 was studied.~\cite{JinDubin,Fine}
The formation of the vortex crystals was predicted by two theories. 
One is a kinetic theory using the Langevin equation.~\cite{Kono}
The other is a theory by the maximizing entropy principle.~\cite{JinDubin2}

Moreover, this phenomenon, namely vortex crystal,
 which is experimentally observed
 was reproduced by numerical simulation.~\cite{Schecter1}
This numerical result showed quantitatively good agreement
 with the experimental one.
Time evolution of the number of intense vortices
 both in experimental flow and in numerical one has similar decay in time.

In the recent experiment,~\cite{Sanpei}
 it is observed that three clumps arrayed
 on a straight line form an equilateral triangular lattice
 under certain conditions of the background.
When the background is very dense, merger immediately occurs instead of 
 crystallization.
In the very dilute background, chaotic motion of the three clumps
 which occurs without the background is not cooled by the background.
Thus the crystallization does not occur and the three clumps
 in the dilute background do not exhibit a triangular configuration.
Therefore the existence and the density of the background
 are supposed to be significant in the process of vortex crystals.

In this paper, we report the results of our studies on these phenomena,
 namely the vortex crystals of three clumps,
 regarded as a $N$-point vortex system.
The vortex crystal was numerically reproduced
 by Schecter \textit{et al}.~\cite{Schecter1}
However their simulation is based on the Euler equation
 (VIC simulation).~\cite{Leonard}
Therefore the details on the dynamics of vortices are not visible,
 namely how vortices in clumps and in background mix each other.
Hence we treat the discrete $N$-point vortex system
 to pursue the behavior of each point vortex. 
In the discrete $N$-point vortex system, the detail of the dynamics
 such as mixing of vortices can be observed.
In order to compare our simulation to the experiments on non-neutral plasma,
 the signs of circulation of vortices are all defined to be positive
 for our $N$-point vortex system in this paper:
 in the experiments on non-neutral plasma,
 the sign of vorticities are all the same.

The $N$-point vortex system is described
 by the following Hamilton equations of motion for open boundary cases,
\begin{equation}
H=-\frac{1}{4\pi}\sum_m\sum_{n\neq m}\Gamma_m\Gamma_n\log | z_m-z_n |,
\end{equation}
\begin{align}
\Gamma_m\frac{\text{d}x_m}{\text{d}t}&=\frac{\partial H}{\partial y_m},&
\Gamma_m\frac{\text{d}y_m}{\text{d}t}&=-\frac{\partial H}{\partial x_m}.
\label{eq:canoHami}
\end{align}
Here $z_m=x_m+\text{i}y_m, \text{i}=\sqrt{-1}$. $(x_m,y_m)$
 and $\Gamma_m$ are the position and the circulation
 of the $m$-th vortex, respectively.
In this case, constants of motion are
 the total circulation $\sum_m \Gamma_m$,
 the angular impulse $I=\sum_m\Gamma_m({x_m}^2+{y_m}^2)$,
 the total energy
 $E=-\frac{1}{4\pi}{\sum\sum}_{(m\neq n)}\Gamma_m\Gamma_n\log |z_m-z_n|$,
 and the center of vorticity $\sum_m \Gamma_m z_m / \sum_m \Gamma_m$.

For the system with circular boundary, the Hamiltonian is given by
\begin{equation}
H =-\frac{1}{4\pi}\sum_m\sum_{n\neq m}\Gamma_m\Gamma_n\log| z_m-z_n| 
+\frac{1}{4\pi}\sum_{m}\sum_{n}\Gamma_m\Gamma_n\log| R^2-z_m\bar{z}_n|
-\frac{\log{R}}{4\pi}(\sum_m \Gamma_m)^2,
\label{eq:hamikabe}
\end{equation}
where $R$ is the radius of the circular boundary. The equations of motion are
\begin{align}
\frac{\text{d}x_m}{\text{d}t}
&=-\sum_{n\neq m}\frac{\Gamma_n}{4\pi}\frac{y_m-y_n}{|z_m-z_n|^2}
+\sum_n
\frac{\Gamma_n}{4\pi}\frac{y_m-{y_n}^{\prime}}{|z_m-{z_n}^{\prime}|^2},
\label{eq:dxkabe}\\
\frac{\text{d}y_m}{\text{d}t}
&=\sum_{n\neq m}\frac{\Gamma_n}{4\pi}\frac{x_m-x_n}{|z_m-z_n|^2}
-\sum_n
\frac{\Gamma_n}{4\pi}\frac{x_m-{x_n}^{\prime}}{|z_m-{z_n}^{\prime}|^2},
\label{eq:dykabe}
\end{align}
where
\begin{align}
{z_m}^{\prime} &= \frac{R^2}{\bar{z}_m}, &
{x_m}^{\prime} &= \frac{R^2}{|z_m|^2}x_m, &
{y_m}^{\prime} &= \frac{R^2}{|z_m|^2}y_m.
\end{align}
The constants of motion for this system with circular boundary are 
the total circulation $\sum_m \Gamma_m$, 
the angular impulse $I$,
and the total energy $E$.

The organization of this paper is as follows.
In a circular domain, a system of
 more than or equal to three point vortices exhibits chaotic behavior.
Therefore, as the most fundamental system of vortices,
 we investigate a three-point vortex system in \S2 for the purpose of
 examining how the system depend on the parameters $(I,E)$,
 namely the angular impulse and the total energy.
By plotting the Poincar\'{e} map,
 we examine the dynamical behavior of this system in detail.
We consider a parameter space which is spanned by two constants of motion
 $(I,E)$.
 For various values of these two parameters,
 the Poincar\'{e} plots are calculated.
We confirm that the topology of the Poincar\'{e} plot
 drastically changes when the values of the parameters are varied
 and that the property of the Poincar\'{e} plot is influenced
 by the sign of ``temperature'' of point vortices
 introduced by L. Onsager.~\cite{Ons}
In \S3, we introduce an expression of the energy spectrum for this $N$-point 
 vortex system with circular boundary.
This energy spectrum is used in order to understand the irreversible merging
 process in which a $N$-point vortex system eventually
 grows into one large vortex,
 which is the result of negative ``temperature''. 
In \S4, we present the results of simulations.
 In a system of discrete point vortices, we reproduce vortex merger
 and a vortex crystal that is observed in the experiments on non-neutral plasma
 and in the simulations based on the 2D Euler equation.
In \S5, we summarize the results.
\section{Poincar\'{e} Section}
In this section, we present the results of the study
  on a three-point vortex system.
We investigate the topology of the Poincar\'{e} plot 
 for the three-point vortex system with 2D circular boundary
 under various initial conditions, which are determined
 by the parameters $(I,E)$, namely the angular impulse and the total energy.
It was confirmed experimentally and 
 will be shown by our numerical simulation in the later section
 that the three clumps system exhibits a vortex crystal
 and vortex merger in the background vorticity distribution.
Studying a three-point vortex system in vacuum is important
 in order to isolate the most fundamental properties from these phenomena,
 i.e., the vortex crystal and vortex merger:
 because it is supposed to be the backbone process
 of the three clumps system in the background vorticity distribution.

It is known that in a system of point vortices without boundary
 chaotic behavior appears for more than or equal to four point vortices,
 and that with circular boundary it appears for more than or equal to three.
Aref and Pomphrey~\cite{Aref} obtained the Poincar\'{e} plot
 for a non-boundary system of four vortices and demonstrated
 that its dynamics is chaotic.
Using their procedure, we calculate the Poincar\'{e}
plots for the dynamics of three point vortices with boundary. 
For the system with circular boundary, there are two constants of 
motion, namely the angular impulse $I$ and the total energy $E$. 
We obtained the Poincar\'{e} plots of this system
 for various values of these two constants.

In the following, according to Aref and Pomphrey~\cite{Aref} 
we derive the Hamiltonian with two degrees  of freedom for our system
by carrying out canonical transformations.
For simplicity, we take the circulation $\Gamma_m=1$ for $m=1,2$ and $3$.
Then the Hamiltonian for this system is
 (The third term of eq.(\ref{eq:hamikabe}) is constant. So we ignore it 
 in the following calculation.)
\begin{align}
H=&-\frac{1}{8\pi}
 \sum_{m=1}^3\sum_{n=1(\neq m)}^3\log(z_m-z_n)(\bar z_m-\bar z_n)
+\frac{1}{8\pi}
\sum_{m=1}^3\sum_{n=1}^3\log(R^2-z_m\bar z_n)(R^2-\bar z_mz_n)\label{eq:Hami}\\
=&\frac{1}{4\pi}\log h(z_1,z_2,z_3), \notag 
\end{align}
where
\begin{align}
h(z_1,z_2,z_3)&=(R^2-|z_1|^2)(R^2-|z_2|^2)(R^2-|z_3|^2)
\frac{|R^2-z_1\bar z_2|^2|R^2-z_2\bar z_3|^2|R^2-z_3\bar z_1|^2}
{|z_1-z_2|^2 |z_2-z_3|^2 |z_3-z_1|^2}.
\end{align}
The vortex coordinates $\{z_i\}_{i=1,2,3}$ are expanded in Fourier series
\begin{equation}
\begin{split}
\begin{cases}
z_1=\frac{1}{\sqrt{3}}
(\sqrt{2J_1}\text{e}^{\text{i}\theta_1}+\sqrt{2J_2}\text{e}^{\text{i}\theta_2}+\sqrt{2J_3}\text{e}^{\text{i}\theta_3}),\\
z_2=\frac{1}{\sqrt{3}}
(\sqrt{2J_1}\text{e}^{\text{i}\theta_1}+\text{e}^{-\text{i}\frac{2\pi}{3}}\sqrt{2J_2}\text{e}^{\text{i}\theta_2}
+\text{e}^{-\text{i}\frac{4\pi}{3}}\sqrt{2J_3}\text{e}^{\text{i}\theta_3}),\\
z_3=\frac{1}{\sqrt{3}}
(\sqrt{2J_1}\text{e}^{\text{i}\theta_1}+\text{e}^{-\text{i}\frac{4\pi}{3}}\sqrt{2J_2}\text{e}^{\text{i}\theta_2}
+\text{e}^{-\text{i}\frac{2\pi}{3}}\sqrt{2J_3}\text{e}^{\text{i}\theta_3}).
\label{eq:z123}
\end{cases}
\end{split}
\end{equation}
Here $J_n$'s and $\theta_n$'s are the action-angle variables.
We further transform the variables in the following way.
\begin{equation}
\begin{cases}
\phi_1=\frac{1}{2}(\theta_1-\theta_3),\\
\phi_2=\frac{1}{2}(\theta_1-2\theta_2+\theta_3),\\
\phi_3=\theta_2,
\end{cases}
\end{equation}
and
\begin{equation}
\begin{cases}
I_1=J_1-J_3,\\
I_2=J_1+J_3,\\
I_3=J_1+J_2+J_3.
\end{cases}
\end{equation}
Substituting them into the Hamiltonian, we get the reduced Hamiltonian 
\begin{equation}
H_{\text{r}}=H_{\text{r}}(I_1,I_2,\phi_1,\phi_2).\label{eq:cHami}
\end{equation}
Here, since $2I_3$ is just the angular impulse (i.e., the constant of motion,
 $I\equiv 2I_3=\sum_{m=1}^3|z_m|^2$),
 then $\phi_3$ is a cyclic coordinate.
Therefore it does not appear in the reduced Hamiltonian.

Furthermore in order to get the Poincar\'{e} section,
 we introduce canonical variables,
\begin{equation}
\begin{cases}
R_1=\sqrt{(I_1+I_3)/2I_3}\cos2\phi_1,\\
P_1=\sqrt{(I_1+I_3)/2I_3}\sin2\phi_1,
\end{cases}
\end{equation}
\begin{equation}
\begin{cases}
{R_2}=\sqrt{(I_3-I_2)/2I_3}\sin2\phi_2,\\
{P_2}=\sqrt{(I_3-I_2)/2I_3}\cos2\phi_2.
\label{eq:RP}
\end{cases}
\end{equation}
The calculation of the Poincar\'{e} plot is done by the following procedure.
 Since the reduced Hamiltonian eq.(\ref{eq:cHami}) is too complicated,
 time evolution is calculated by using the Hamiltonian eq.(\ref{eq:Hami}). 
We use the 4th order Runge-Kutta method
 for our numerical calculation. 10000 random initial points are given on
 the $(R_1,P_1)$-plane satisfying two constants of motion,
 namely the angular impulse and the total energy, and time evolution is
 done until these each orbit of initial points crosses the $(R_1,P_1)$-plane
 50 times. The Poincar\'{e} section is set on the plane with
 $R_2=0, \dot{R_2}>0$.
In this simulation, relative errors of both constants of motion,
 namely $I$ and $E$, are $10^{-10}$.

\begin{figure}[p]
\begin{center}
\rotatebox{-90}{\includegraphics[width=10.0cm]{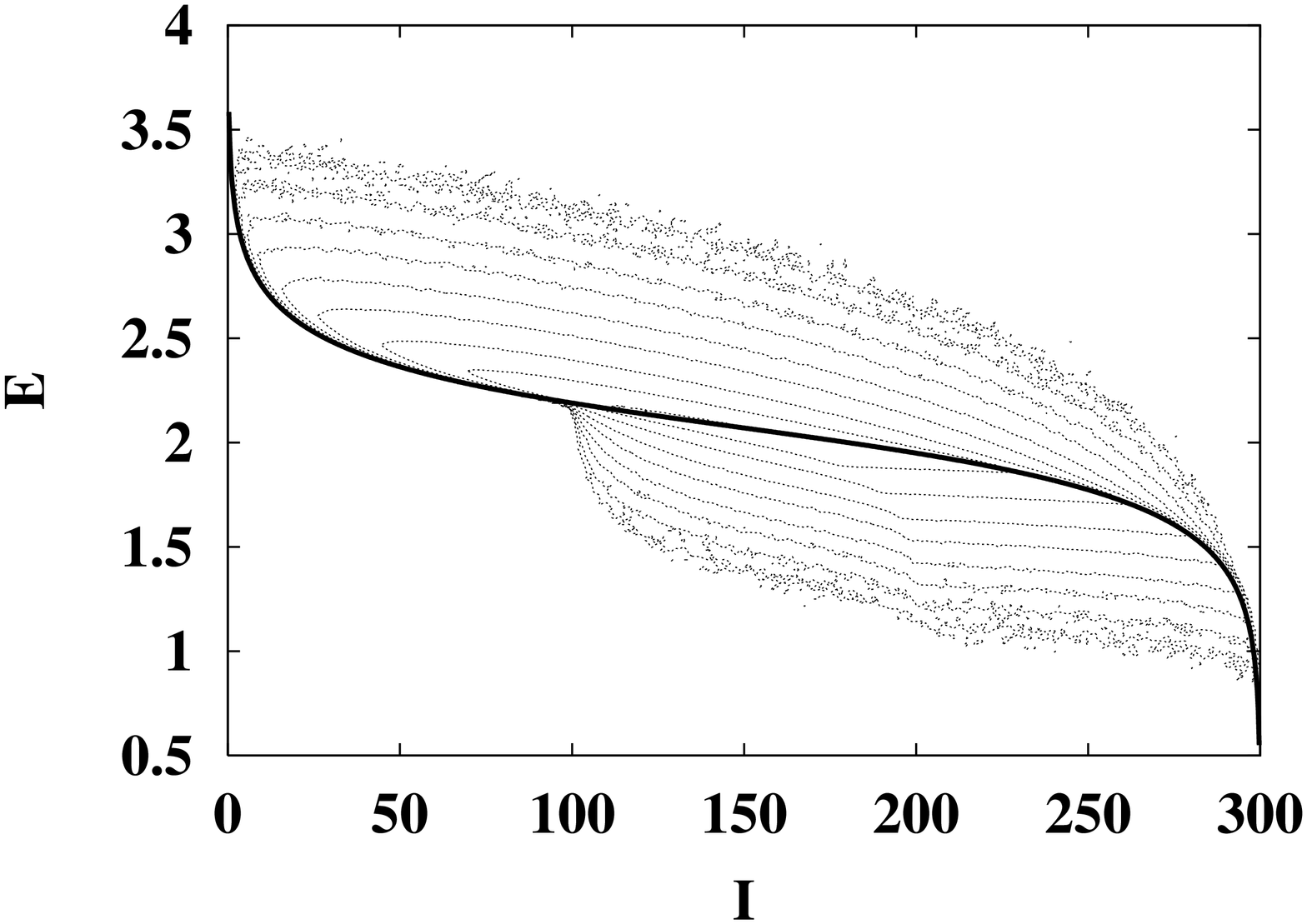}}
\end{center}
\caption{
The contour plot of the density of states as a function of $I$ and
 $E$ calculated from $5.0 \times 10^7$ random configurations of three
  vortices.
 A rigid line in this picture shows the values of parameters at equilateral
 triangle configuration of which center of mass is at the origin.
 The dotted lines indicate the contour levels 
 (these are from outer to inner, $1.0\times 10^{-5},3.0\times 10^{-5},1.0\times 10^{-4},3.0\times 10^{-4},1.0\times 10^{-3},3.0\times 10^{-3},1.0\times 10^{-2},3.0\times 10^{-2},1.0\times 10^{-1}$(\%) configurations of all, respectivelly.).
}
\label{fig:DHI}
\end{figure}

First, we show the density of states as a function of the angular
 impulse $I$ and the total energy $E$ as Johnson did
 (Fig.\ref{fig:DHI}).~\cite{Darell}
A rigid line represents the values of the parameters
 at which three point vortices form an equilateral triangle configuration;
 the center of mass of the equilateral triangle is at the origin.
The form of the rigid line is given by
\begin{equation}
E_{\text{r}}(I)=\frac{1}{4\pi}
\log \left( \frac{1}{27}\frac{[R^6-(I/3)^3]^3}{(I/3)^3}\right).
\label{eq:regtri}
\end{equation}
Here $R$ is the radius of the circular boundary.
The dotted lines indicate several contour levels of density of states.
The innermost level is the highest (from outer to inner,
 $1.0\times 10^{-5},3.0\times 10^{-5},1.0\times 10^{-4},
3.0\times 10^{-4},1.0\times 10^{-3},3.0\times 10^{-3},
1.0\times 10^{-2},3.0\times 10^{-2},1.0\times 10^{-1}$(\%) 
 of all configurations, respectively).

In the region for small $I$ in Fig.\ref{fig:DHI},
 the states with $E<E_{\text{r}}$ do not exist.
 The reason why the density in this region of parameters is zero 
 is as follows.
When vortices are close each other, the energy $E$ increases.
If vortices are close to boundary, the energy decreases. 
In small $I$ range, because of the restriction of the angular impulse $I$, 
 no vortices can be close to the boundary.
Therefore, in the region of small $I$,
 the equilateral triangular configuration (rigid line)
 gives the minimum energy. This is actually observed in Fig.\ref{fig:DHI}.

\begin{figure}[p]
  \begin{center}
\includegraphics{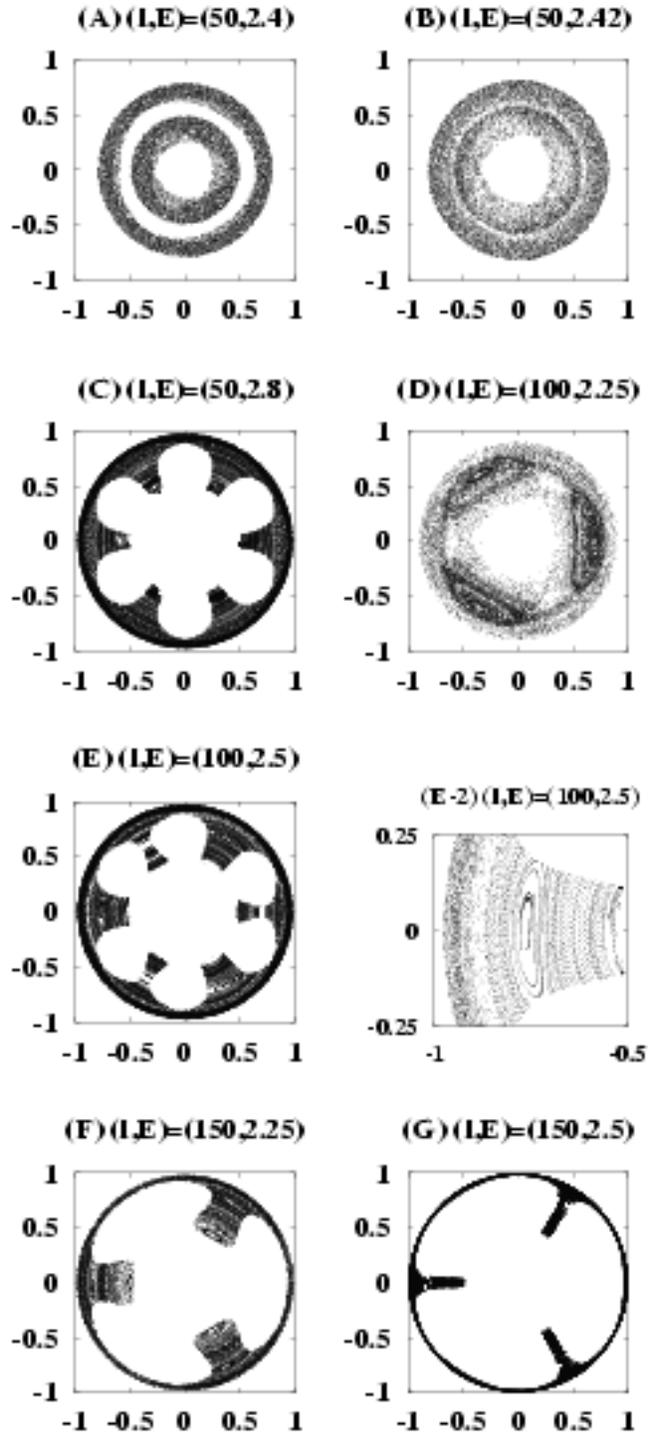}
  \end{center}
\caption{
The Poincar\'{e} plots on the $(P_1,R_1)$-plane
 for various values of parameters. 
 The values of parameters are summarized in Table \ref{tbl:vl-PS}.
(E-2) is the enlargement of (E).
}
\label{Fig:PS}
\end{figure}

\begin{figure}[p]
\begin{center}
{\includegraphics[width=15.0cm]{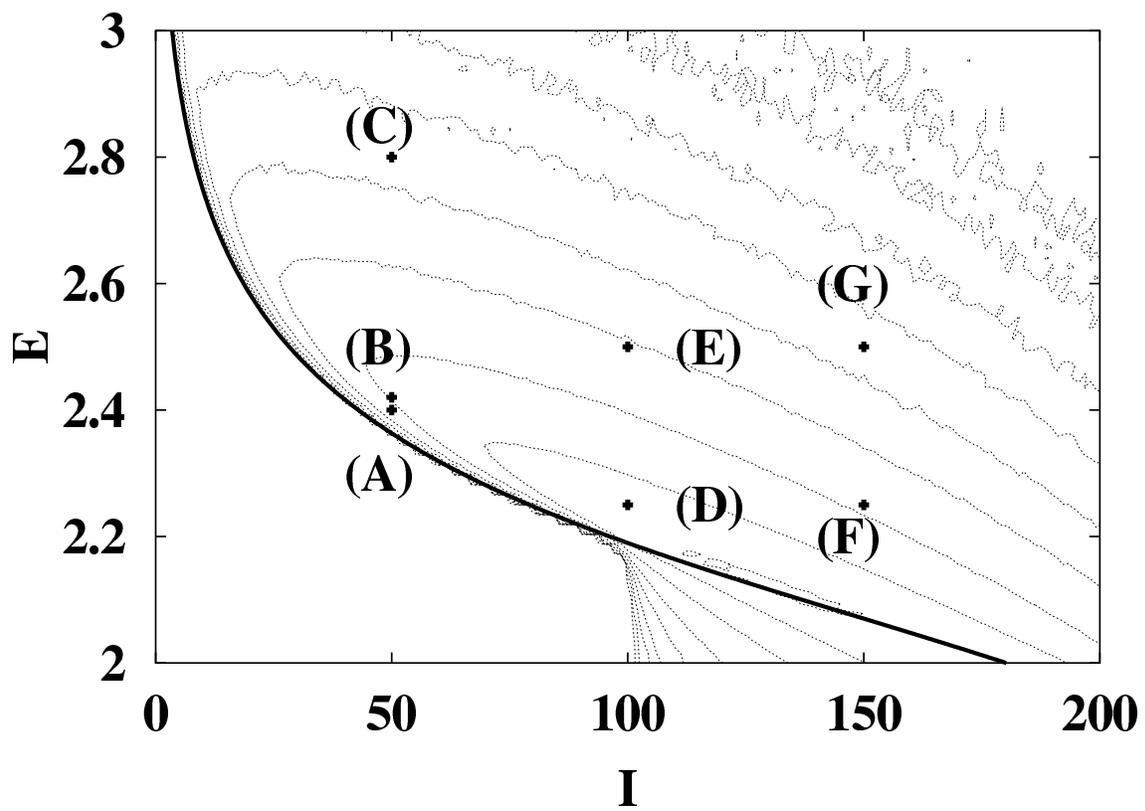}}
\end{center}
\caption{
The function $E_{\text{r}}(I)$ in which three vortices form
 an equilateral triangle configuration (rigid line)
 and the values of the parameters for Poincar\'{e} plots
 in Figs.\ref{Fig:PS}(A)-\ref{Fig:PS}(G).
 (A)-(G) correspond to the sections (A)-(G) in Fig.\ref{Fig:PS}, respectively.
}
\label{Fig:souzu}
\end{figure}

Figure \ref{Fig:PS} shows the Poincar\'{e} plots for some parameters.
The values of these parameters are indicated in Fig.\ref{Fig:souzu}
 and Table \ref{tbl:vl-PS}.
The density of the states reflects the area of the plots on the
 Poincar\'{e} section.
For example, we compare the areas of Fig.\ref{Fig:PS}(F)
 with that of Fig.\ref{Fig:PS}(G).
Here we express the region of the plots in Fig.\ref{Fig:PS}(F)
 as $\mathcal{D}_{\text{(F)}}$ on the complex plane $Z=P_1+\sqrt{-1} R_1$.
Thus the region $\mathcal{D}_{\text{(F)}}$ is expressed as
 $\mathcal{D}_{\text{(F)}}
={\mathcal{D}_{\text{(F)}}}^{(0)}
\bigcup{\mathcal{D}_{\text{(F)}}}^{(1)}
\bigcup{\mathcal{D}_{\text{(F)}}}^{(2)}
\bigcup{\mathcal{D}_{\text{(F)}}}^{(3)}$,
 where ${\mathcal{D}_{\text{(F)}}}^{(0)}=\{Z|0<s<|Z|<l\}$, and
 ${\mathcal{D}_{\text{(F)}}}^{(1)}$, ${\mathcal{D}_{\text{(F)}}}^{(2)}$ and 
 ${\mathcal{D}_{\text{(F)}}}^{(3)}$ represent three protuberances sprouted
 from ${\mathcal{D}_{\text{(F)}}}^{(0)}$ toward the origin.
As the energy increases, three regions of the protuberances become thinner
 along the azimuthal direction. The region ${\mathcal{D}_{\text{(F)}}}^{(0)}$
 also becomes thinner (Fig.\ref{Fig:PS}(G)).
This tendency is observed for general values of $I$.

When $E=E_{\text{r}}$ (hence the parameters $(I,E)$ are on the rigid line
 in Fig.\ref{fig:DHI}), we can see from eqs.(\ref{eq:z123}-\ref{eq:RP})
 that the plots on the Poincar\'{e} section 
 are lying on the circle of which radius is $\sqrt{1/2}$ and at the origin.
When the value of $E$ changes from $E_{\text{r}}$ slightly for the same value of $I$,
 the region of the plots becomes concentric rings as Fig.\ref{Fig:PS}(A):
 if we write this two regions on the complex plane $Z=P_1 + \sqrt{-1}R_1$
 as $\mathcal{D}_1$ and $\mathcal{D}_2$, when $E=E_{\text{r}}$ these two region
 is expressed as $\mathcal{D}_1=\{Z| |Z|=0\}$ and
 $\mathcal{D}_2=\{Z| |Z|=\sqrt{1/2}\}$.
When $E$ changes from $E_{\text{r}}$, these regions change to
 $\mathcal{D}_1=\{Z|a_1\leq |Z|\leq a_2\}$ and
 $\mathcal{D}_2=\{Z|b_1\leq |Z|\leq b_2\}$, where $0<a_1<a_2<b_1<b_2$.
Therefore, the orbit in the phase space
 do not pass the origin on the $(P_1,R_1)$-plane.
Actually, in Fig.\ref{Fig:PS}, there are no points
 around the origin on the Poincar\'{e} section.
This means that the topology of the Poincar\'{e} plot changes
 when the value of $E$ crosses $E_{\text{r}}$. 
As $E$ increases, this two regions of the concentric rings
 are connected as seen in Fig.\ref{Fig:PS}(B):
 these two regions $\mathcal{D}_1$ and $\mathcal{D}_2$ defined above 
 change to  $\mathcal{D}_1=\{Z|a_1\leq |Z|\leq a_2\}$ and
 $\mathcal{D}_2=\{Z|b_1\leq |Z|\leq b_2\}$, where $0<a_1<b_1<a_2<b_2$.
As $E$ increases more, the region $\mathcal{D}_1\bigcup\mathcal{D}_2$ 
 is scooped out:
 $\mathcal{D}_1\bigcup\mathcal{D}_2\setminus\bigcup_{i=1}^m\mathcal{D}^{(i)}$,
 where $\mathcal{D}^{(i)}$ are the scooped region
 as seen in Fig.\ref{Fig:PS}(C) (for m=6) and Fig. \ref{Fig:PS}(G) (for m=3).

Furthermore, the property of Poincar\'{e} plot is apparently changed
 when the sign of the gradient of the density of states
 $\partial{W}/\partial{E}$ is changed to opposite sign.
 This gradient is proportional to the inverse temperature $1/T$
 defined by ``statistical mechanics''.
\begin{equation}
\begin{split}
S&=\log W,  \\
\frac{1}{T}
&=\frac{\partial{S}}{\partial{E}}
=\frac{1}{W} \frac{\partial{W}}{\partial{E}},
\end{split}
\end{equation}
where $S$ is the entropy and $E$ is the energy of the system.
 See Fig.\ref{Fig:souzu}.
 The values of the energy of points (A) and (B)
 are lower than the ridge of contour line
 (positive ``temperature''), and (D) is around the ridge
 (``temperature'' is infinite).
 The others are on the region upper than the ridge (negative ``temperature'').
 When the ``temperature'' is negative, the region of Poincar\'{e} plot
 is scooped out.
This shows that the sign of ``temperature'' influences the behavior
 of the Poincar\'{e} plot and the phase space structure.

\begin{figure}[p]
  \begin{center}
\includegraphics{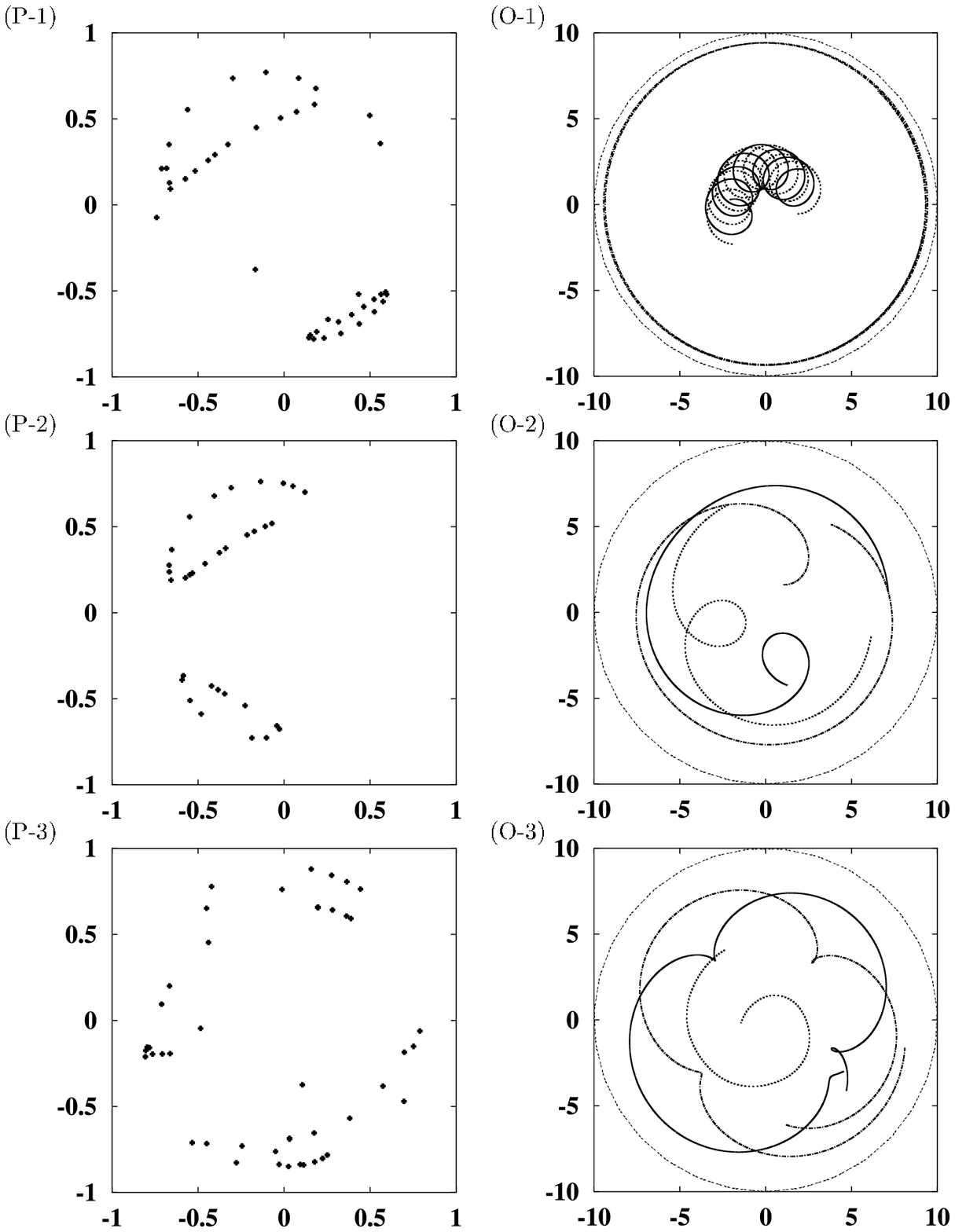}
  \end{center}
\caption{
The Poincar\'{e} plot of a single orbit in phase space (P-1,2,3)
 and the locus of vortices on real plane (O-1,2,3)
 under different initial condition with $(I,E)=(100,2.25)$.
 (P-1,2,3) are the parts of Fig.\ref{Fig:PS}(D).
}
\label{Fig:PS-VC}
\end{figure}

In the following, we take notice of the properties of the motion
 in the three-point vortex system with the parameter $(I,E)=(100,2.25)$.
Figure \ref{Fig:PS-VC} shows the Poincar\'{e} plot and the locus of three point
 vortices under different initial conditions on the real $(x,y)$-plane
 with $(I,E)=(100,2.25)$.
The plots of Figs.\ref{Fig:PS-VC}(P-1) - \ref{Fig:PS-VC}(P-3)
 are generated from a single trajectory under different initial conditions.
Figs.\ref{Fig:PS-VC}(O-1) - \ref{Fig:PS-VC}(O-3)
 show the motions of rotating three vortices.
In the case of Fig.\ref{Fig:PS-VC}(O-1),
 two vortices rotate each other around the origin
 apart from the other vortex which runs near the boundary. 
We will call the motion of this two vortices rotating each other
 in Fig.\ref{Fig:PS-VC}(O-1) the ``binary star motion''.
As seen in Fig.\ref{Fig:PS-VC}(P-1), this motion of voritices plots
 tori on the $(P_1,R_1)$-plane.
Tori on the Poincar\'{e} section which does not correspond to
 the ``binary star motion'' of vortices are shown in Fig.\ref{Fig:PS-VC}(P-2).
The corresponding locus is depicted in Fig.\ref{Fig:PS-VC}(O-2).
Though the loci of these two type of motions
 in Figs.\ref{Fig:PS-VC}(O-1) and \ref{Fig:PS-VC}(O-2)
 are apparently different, these Poincar\'{e} plots in 
 Figs.\ref{Fig:PS-VC}(P-1) and \ref{Fig:PS-VC}(P-2)
 are similar in its regular behavior on the Poincar\'{e} section
 (Figs.\ref{Fig:PS-VC}(P-1) and \ref{Fig:PS-VC}(P-2)).
The other type of motions of vortices is shown in Figs.\ref{Fig:PS-VC}(P-3)
 and \ref{Fig:PS-VC}(O-3). In this case, ``chaotic sea'' emerges
 on the $(P_1,R_1)$-plane in Fig.\ref{Fig:PS-VC}(P-3).
 The vortex orbits show cusps as in Fig.\ref{Fig:PS-VC}(O-3).
As mentioned above,
 the motion of the three-point vortex system can be classified 
 at least three types, namely the types
 of Figs.\ref{Fig:PS-VC}(O-1) - \ref{Fig:PS-VC}(O-3).
The behavior of the Poincar\'{e} plot for the motion of vortices enable us
 to distinguish the properties of vortices motion
 as in Figs.\ref{Fig:PS-VC}(P-2) and \ref{Fig:PS-VC}(P-3):
 this difference is not distinguishable
 by only seeing the loci of three vortices
 in Figs.\ref{Fig:PS-VC}(O-2) and \ref{Fig:PS-VC}(O-3).

It is difficult to calculate the Poincar\'{e} plot
 for the parameters in the region $E<E_{\text{r}}$ and in the region of large $I$. 
We suppose that there are two reasons why this calculation is difficult. 
The first reason is the complexity of the formula
 of the reduced Hamiltonian eq.(\ref{eq:cHami}).
This complexity causes numerical errors. 
The second reason is the following.
When $E<E_{\text{r}}$, a vortex runs quite near the boundary (hence it runs very fast). 
In this situation we have to take the time step to be small. 
In actual numerical computation we fixed the time step. 
Therefore in that case numerical error becomes large
 in the calculation of the Poincar\'{e} plot.
This difficulty of numerical integration is a subject to be solved.
A variable time step scheme would be needed.
\begin{table}[p]
\begin{center}
	\begin{tabular}{lcrc}					\hline
	$I$	&$E_{\text{r}}$	&	$E$	&	index	\\ \hline
	50	&2.361	&	2.4	&	(A)	\\
		&	&	2.42	&	(B)	\\
		&	&	2.8	&	(C)	\\
	100	&2.189	&	2.202	&	(D)	\\
		&	&	2.5	&	(E)	\\
	150	&2.070	&	2.25	&	(F)	\\
		&	&	2.5	&	(G)	\\ \hline
	\end{tabular}
	\caption{
The values of the parameters $(I,E)$ for the Poincar\'{e} plots
 shown in Fig.\ref{Fig:PS}. $E_{\text{r}}$ indicates
 the value of the energy for the equilateral triangle configuration.
}
\label{tbl:vl-PS}
\end{center}
\end{table}
\section{Energy Spectrum}
In a $N$-point vortex system,
 when the temperature is negative,
 many like-signed vortices 
 merge each other and tend to become a large vortex.~\cite{Ons}
 This indicates that the energy is transferred from small scale
 to large scale in two dimensional case.
This phenomenon is known
 as the inverse cascade~\cite{Ranryuu0,Ranryuu1,Ranryuu2}
 in general two dimensional turbulence.

Novikov gave a formula for the energy spectrum
 for a system of point vortices without boundary.~\cite{Novikov}
 In the following, according to his procedure,
 we derive an explicit expression of the energy spectrum
 for a $N$-point vortex system with circular boundary.

The energy is defined by
\begin{equation}
E=\int_0^{\infty}\tilde{E}(k)\text{d} k
=\int_0^{\infty}\frac{| \tilde{\bm v}(\bm k)|^2}{2} \text{d}k
=\int_0^{\infty}\frac{| \tilde{\omega}(\bm k)|^2}{2k^2} \text{d}k.
\end{equation}
Here, $\tilde{\bm{v}}(\bm{k})$ is the Fourier transform of velocity field,
\begin{equation}
{\tilde{\bm{v}}}(\bm{k})
=\frac{1}{2\pi}\int_{-\infty}^{\infty}
\bm{v(r)}\exp(\text{i}\bm{k}\cdot\bm{r})\text{d}\bm{r},
\end{equation}
and $\tilde{\omega}(\bm{k})$ is the Fourier transform of the vorticity field.
In this system, the velocity field is restricted
 to be within the circular domain. Thus we introduce an induced vortex sheet
 on the circular boundary whose radius is R. The vorticity field is given by
\begin{align}
\omega(\bm{r})
&=\sum_m\Gamma_m\delta(\bm{r}-\bm{r}_m)
-\left[\nabla\Psi\cdot\bm{n}\right]_R\delta(r-R)  \\
&=\sum_m\Gamma_m\delta(\bm{r}-\bm{r}_m)
-v_\theta |_R  \delta(r-R) \label{eq:IdcVor} \\
&=\sum_m\Gamma_m\left[\delta(\bm{r}-\bm{r}_m)
-\frac{R^2-{r_m}^2}{2\pi R | \bm{R}-\bm{r}_m|^2}
\delta(r-R) \right].\label{eq:IdcVor2}
\end{align}
The effect of the induced vortex sheet is
 given by the second term in each line.
Here $\bm{r}=(x,y)$, $\bm{r}_m$ is the position of the $m$-th vortex,
 and $\bm{R}=R\bm{r}/r$.
 $\Psi(\bm{r})$ is the stream function by vortices
 which is within the boundary,
 $\bm{n}$ is a unit vector perpendicular to the boundary,
 and $\nabla{\Psi}\cdot\bm{n}=-\partial\Psi/\partial r=v_{\theta}$.
 The derivation of the second term in eq.(\ref{eq:IdcVor2})
 is given in Appendix A.

Thus the Fourier transform of the vorticity field $\omega(\bm{r})$
 is given by,
\begin{align}
\tilde\omega(\bm{k})
&=\frac{1}{2\pi}
\int\omega(\bm{r})\exp(\text{i}\bm{k}\cdot\bm{r})\text{d}\bm{r} \\
&=\frac{1}{2\pi}\int\sum_m\Gamma_m\left[\delta(\bm{r}-\bm{r}_m)
-\frac{R^2-{r_m}^2}{2\pi R | \bm{R-r_m}|^2}\delta(r-R) \right]
\exp(\text{i}\bm{k}\cdot\bm{r})\text{d}\bm{r}.
\end{align}
This can be easily rewritten into 
\begin{align}
\tilde\omega(\bm{k})
&=\frac{1}{2\pi}\sum_m \Gamma_m
\left[
 \int\delta(\bm{r-r_m})\exp(\text{i}\bm{k}\cdot\bm{r})\text{d}\bm{r}
 -\frac{R^2-{r_m}^2}{2\pi R} \oint_{r=R}
\frac{\exp(\text{i}\bm{k}\cdot\bm{R})}{| \bm{R}-\bm{r}_m|^2}\text{d}\bm{r}
\right]\\
&=\frac{1}{2\pi}\sum_m \Gamma_m\left[
1
-\frac{R^2-{r_m}^2}{2\pi R}
\oint_{r=R} 
\frac{\exp[\text{i}\bm{k}\cdot(\bm{R}-\bm{r}_m)]}
{| \bm{R}-\bm{r}_m|^2} \text{d}\bm{r} \right]
\exp(\text{i}\bm{k}\cdot\bm{r}_m)    \\
&=\sum_m\tilde{\Gamma}_m(\bm{k})\exp(\text{i}\bm{k}\cdot\bm{r}_m),
\end{align}
where we define
\begin{equation}
\tilde{\Gamma}_m(\bm{k}) \equiv \frac{1}{2\pi}\Gamma_m\left[
1
-\frac{R^2-{r_m}^2}{2\pi R}
\oint_{r=R} 
\frac{\exp[\text{i}\bm{k}\cdot(\bm{R}-\bm{r}_m)]}
{| \bm{R}-\bm{r}_m|^2} \text{d}\bm{r}
\right].
\end{equation}
In the limit $k\to 0$, $\tilde{\Gamma}_m(\bm{k})$ is vanished.
Actually, the vorticity disappears in the outside of the circular domain.

By definition,
 $\overline{\tilde{\Gamma}_m(\bm{k})}=\tilde{\Gamma}_m(-\bm{k}) $.
Thus we have  
\begin{equation}
| \tilde{\bm v}(\bm k)|^2 = \frac{| \tilde\omega(\bm{k})|^2}{k^2}
=\frac{1}{k^2}
\sum_m\sum_n\tilde{\Gamma}_m(\bm{k})\tilde{\Gamma}_n(-\bm{k})
\exp[\text{i}\bm{k}\cdot(\bm{r}_m-\bm{r}_n)].
\end{equation}
After some calculation, we have the formula for the energy spectrum,
\begin{equation}
\begin{split}
\tilde{E}(k)=&\frac{1}{4\pi k} \left\{ 
\sum_n{\Gamma_n}^2  \right.  
+2\sum_m\sum_{n<m}\Gamma_m\Gamma_nJ_0(k| \bm{r}_m-\bm{r}_n|) \\
&\qquad -2\sum_m\sum_n\Gamma_m\Gamma_n
 \sum_{l=0}^{\infty}\varepsilon_l\left(\frac{r_m}{R}\right)^lJ_l(kR)J_l(kr_n)
 \cos[l(\varphi_m-\varphi_n)] \\
&\qquad +\sum_m\sum_n\Gamma_m\Gamma_n
\sum_{l=0}^{\infty} \varepsilon_l
\left[J_l(kR)\right]^2
\left. \left(\frac{r_mr_n}{R^2}\right)^l\cos[l(\varphi_m-\varphi_n)]   \right\},
\label{eq:ES}
\end{split}
\end{equation}
where $\varphi_m$ is the angle between $\bm{r}_m$ and the zero axis,
 and $z_m = r_m \text{e}^{\text{i}\varphi_m}$.
 $J_l(x)$ is the $l$-th Bessel function. $\varepsilon_l$ is defined as
\begin{equation}
\varepsilon_l= \begin{cases}
	1 & (l=0) \\
	2 & (l\geq1).
	\end{cases} 
\end{equation}
Detail on the derivation is shown in Appendix B.
Note that the first two terms in eq.(\ref{eq:ES}) are just the formula derived
 by Novikov~\cite{Novikov} and then the rest terms represent the effect of the 
 circular boundary.
\section{Numerical Simulation}
In this section, we show the results of our numerical simulations
 on a vortex crystal and vortex merger. 
We confirm that the system of a few hundreds discrete point vortices
 is also able to exhibit a vortex crystal and a vortex merger.

Non-neutral plasma experiments and 2D Euler equation numerical simulations
 found that by the existence of continuous
 low background vorticity distribution,
 in certain initial conditions intense vortices (clumps)
 are arrayed in a triangular lattice (vortex crystal).
When the initial condition changes slightly to other distribution, 
 clumps merge each other (vortex merger).
This merger is the phenomenon seen when the ``temperature'' is negative.

In our numerical calculations,
 the Hamilton equations of the $N$-point vortex system
 are numerically integrated, not using the Euler equation. 
Numerical simulation is performed in the following procedure.
We numerically integrate the equations of motion
 of point vortices with circular boundary
 (whose radius R=10), namely eqs.(\ref{eq:dxkabe}) and (\ref{eq:dykabe}),
  by the 4th-order Runge-Kutta method. 
%
%
In our simulation, a relative error of the constants of motion
 is $10^{-7}-10^{-8}$ for energy $E$ and $10^{-10}-10^{-11}$
 for angular impulse $I$.
In the simulation using the Euler equation
 by Schecter \textit{et al}.,~\cite{Schecter1}
 errors of both constants of motion are about $10^{-3}$.
\subsection{Vortex crystals}
In the experiment by Sanpei \textit{et al}.,~\cite{Sanpei}
 it is found that three clumps are
 accelerated  to form an equilateral triangular configuration
 by interaction with continuous low-level background vorticity distribution.
Since three clumps make a unit cell of vortex crystals, it is regarded
 that this process is a fundamental process of forming vortex crystals.
The existence of continuous low-level background vorticity distribution
 and the interaction between the background and the clumps
 are regarded as a key factor for making vortex crystals.

\begin{figure}[p]
  \begin{center}
\includegraphics{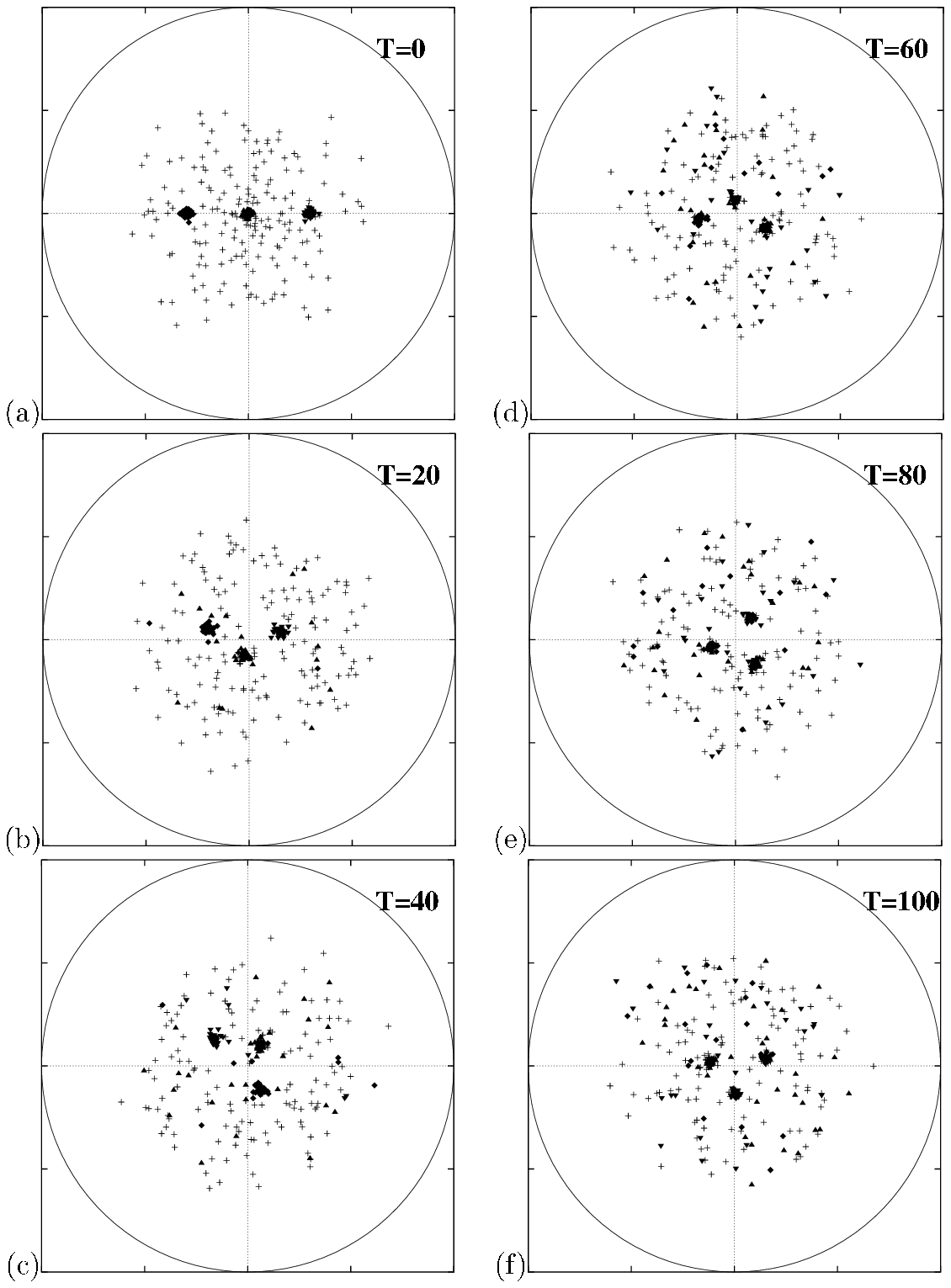}
\caption{
The vortex crystal of point vortices. Each clump contains 60 vortices.
 The background consists of 180 vortices.
 The circulation of each point vortex is 0.2.
}
   \label{Fig:VrCry}
  \end{center}
\end{figure}

We reproduce qualitatively a vortex crystal of three clumps
 in a system of point vortices.
In Fig.\ref{Fig:VrCry}, our result is depicted.
In our simulation, for the initial condition, 
 each clump (densely distributed regions of point vortices)
 contains about 60 vortices, and the background distribution consist of
 about 180 vortices. The circulation of each vortex is 0.2.
Three Clumps are arrayed on a straight-line with one clump at equal distances
 from others (Fig.\ref{Fig:VrCry}(a)) so as to be the same as the experiment
 by Sanpei \textit{et al}.~\cite{Sanpei}
The middle clump contains more point vortices than others
 because of the overlap with background point vortices.

Before seeing the result of  vortex crystallization of our simulation,
 we define the property of clumps configuration, namely ``arranging order''.
Labeling each clump by $C_i (i=1,2$ and $3$), rotating sequence of clumps
 in anti-clockwise is only $(C_1C_2C_3)$ or $(C_3C_2C_1)$.
We call these rotating sequence the ``arranging order''.
Vortex crystallization occurs as the following.
At early steps,
 the clumps are at vertices of non-equilateral triangle rotating
 in the background point vortices
 (in Figs.\ref{Fig:VrCry}(a)-\ref{Fig:VrCry}(c)). 
Although the arranging order of the clumps do not change,
 the side length of the triangle of clumps changes.
At later steps,
 the clumps form equilateral triangle and these rotate quasi-stationarily
 for a long time (we calculated until $T=200$)
 (in Figs.\ref{Fig:VrCry}(d)-\ref{Fig:VrCry}(f)).

In addition to the ``arranging order'',
 we introduce a signed symmetry parameter $S$. 
A symmetry parameter is defined
 as $S=12\sqrt{3}A/l^2$ by Sanpei \textit{et al}.\cite{Sanpei}
$A$ is the area and $l$ is the peripheral length of
 the triangle whose vertices are the clumps.
We define $A$ as $\bm{e}_z\cdot(\bm{r}_1-\bm{r}_3)\times(\bm{r}_2-\bm{r}_3)/2$.
 Here $\bm{e}_z$ is a unit vector perpendicular to $(x,y)$-plane,
 and $\bm{r}_i$ is the position of the $i$-th clump.
An absolute value of $S$ is maximized at 1
 when the clumps form an equilateral triangle,
 and the sign of $S$ is changed when the ``arranging order'' changes.
In the following result of our simulation, the sign of $S$ does not change
 because the ``arranging order'' does not change.

\begin{figure}[p]
\begin{center}
\includegraphics[width=15.0cm]{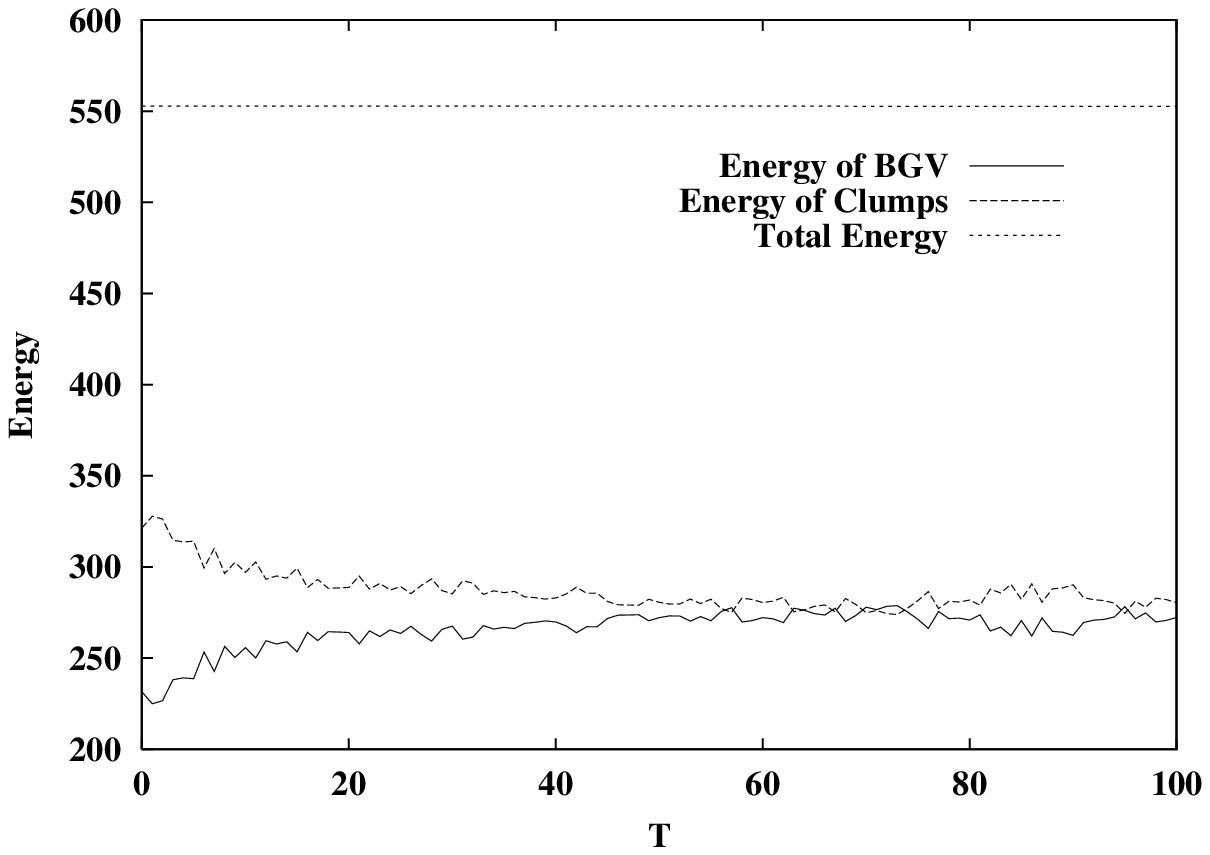}
\end{center}
\caption{
The energy of three clumps, the energy of the background vortices
 and the total energy of the system. The energy of the clumps tends to transfer
 toward the background. In later steps (the time crystallization occur), 
 the energy transfer saturates.
}
\label{Fig:EEE}
\end{figure}

\begin{figure}[p]
\begin{center}
\includegraphics[width=15.0cm]{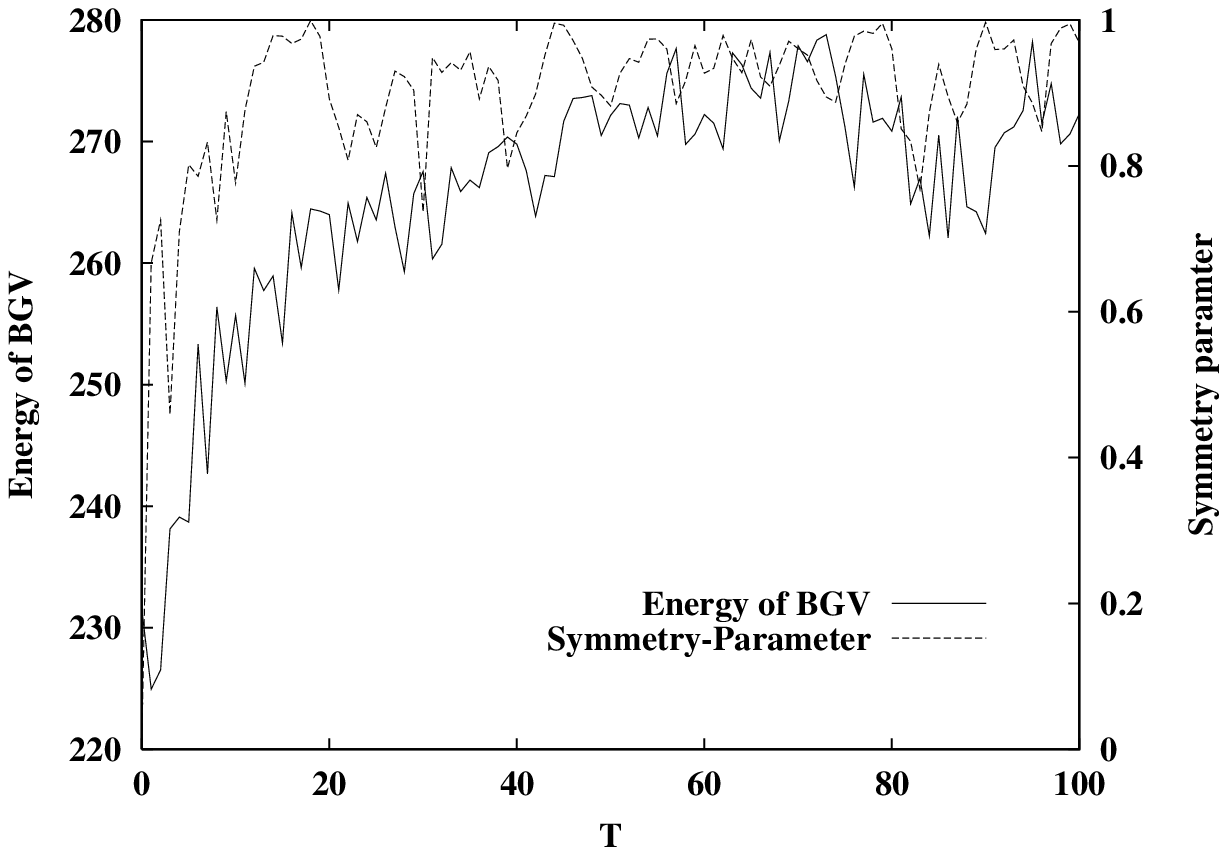}
\end{center}
\caption{
The signed symmetry parameter and the energy of the background vorticity.
These show similar tendency that when the clumps form triangular lattice
 these value are growing.
}
\label{Fig:S-E}
\end{figure}

We confirmed that the energy of clumps is transferred to the background
 vortices in the process of vortex crystallization (Fig.\ref{Fig:EEE}).
The energy of three clumps decreases and the energy of the background increases
 while total energy of the system is conserved.
In this process, the tendency of the energy of the background vortices is
 similar to that of the signed symmetry parameter(Fig.\ref{Fig:S-E}).
%
%
%
Therefore, this means that if there is an outlet of energy of clumps,
 i.e., the background, three clumps tend to form a triangular lattice.
However, it is necessary for the clumps not to occur merging.

In this simulation, the arranging order of the clumps
 (namely, the sign of the signed symmetry parameter $S$) does not change. 
However, in several parameters (the number of vortices,
 the circulation of background or clumps, and the ratio of these)
 this arranging order does change,
 i.e., the system shows the ``binary star motion'', 
 and the system does not exhibit vortex crystals.
In ref.~\citen{Schecter1}, it is noticed 
 that the ratio of the circulation of background to that of clumps
 influences the relaxation rate toward vortex crystals. 
According to the simulation by Schecter et. al,~\cite{Schecter1}
 the relaxation rate toward vortex crystals
 is the fastest when $\Gamma_{\text{background}}/\Gamma_{\text{total}}$
 is $0.2-0.4$.
In our case, $\Gamma_{\text{background}}/\Gamma_{\text{total}}=0.5$.
We suppose that the background distribution
 with the proper ratio to total circulation
 restrains the ``binary star motion'' to crystallize the clumps.

It is observed that the point vortices of the clumps and
 the background mix each other (Fig.\ref{Fig:VrCry}).
 However, the number of vortices of each clump is conserved around 60.
The vortices of the clumps indicated by $\blacktriangle$,
 $\blacktriangledown$ and $\blacklozenge$ are oozed out to the background.
The vortices of the background indicated by $+$ permeate to the clumps.
\subsection{Merging}
The background vortices causes not only crystallization but also merging.
In this subsection, we consider the $N$-point vortex system
 as a model of vortex merging.
Merging is highly important in a process of decay into vortex crystals,
 or into equilibrium states.
Furthermore the power law behavior of the number of clumps
 ($N\sim t^{-\xi}, \xi=0.2-0.7$)~\cite{Fine,JinDubin}
 in vortex crystals occur through the merging process of clumps.
In merging process, we also calculate the energy spectrum of the system.
In our simulation, each clump and the background contain $30$ point vortices.
The circulation $\Gamma_m$ is set as $\Gamma_m=1$.
For the initial configuration, the clumps are set at the vertices
 of equilateral triangle whose center of mass is at the origin.
The energy spectrum is calculated
 by the formula derived in the previous section. 

\begin{figure}[p]	
  \begin{center}
\includegraphics{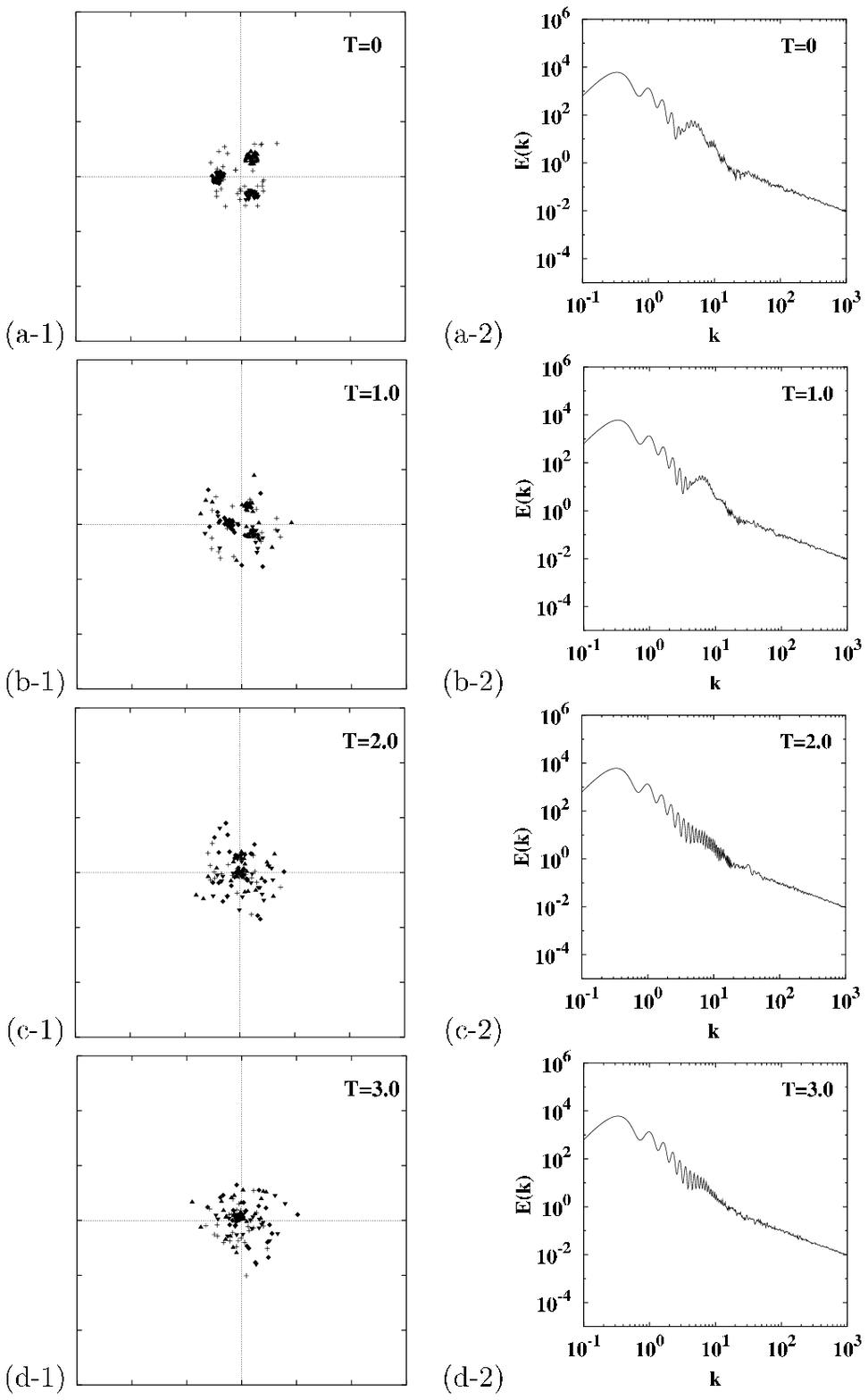}
    \caption{
The configurations of vortices and the energy spectrum in the merging process.
Each clump and the background consist of 30 point vortices.
Around $T=2.0$, a vortex merger occurs.
}
\label{Fig:VrMg}
  \end{center}
\end{figure}

We show the configuration of point vortices
 and the energy spectrum of this system
 in Fig.\ref{Fig:VrMg}. The circular boundary
 is out of view (the area shown in Fig.\ref{Fig:VrMg} is the region
 $[-6,6]\times[-6,6]$, the radius of boundary is $R=10$).
The behavior of peaks of the energy spectrum
 of the range $10^0<k<10^1$ drastically changes in merging process.
Before merging (in a lattice state, Figs.\ref{Fig:VrMg}(a)
 and \ref{Fig:VrMg}(b)), the peaks of the energy spectrum
 are oscillated with relatively large period in the range $10^0<k<10^1$.
However, after merger (Figs.\ref{Fig:VrMg}(c) and \ref{Fig:VrMg}(d))
 these become very sharp and wildly oscillates
 (i.e., plenty of peaks).

\begin{figure}
\begin{center}
\includegraphics[width=10.0cm]{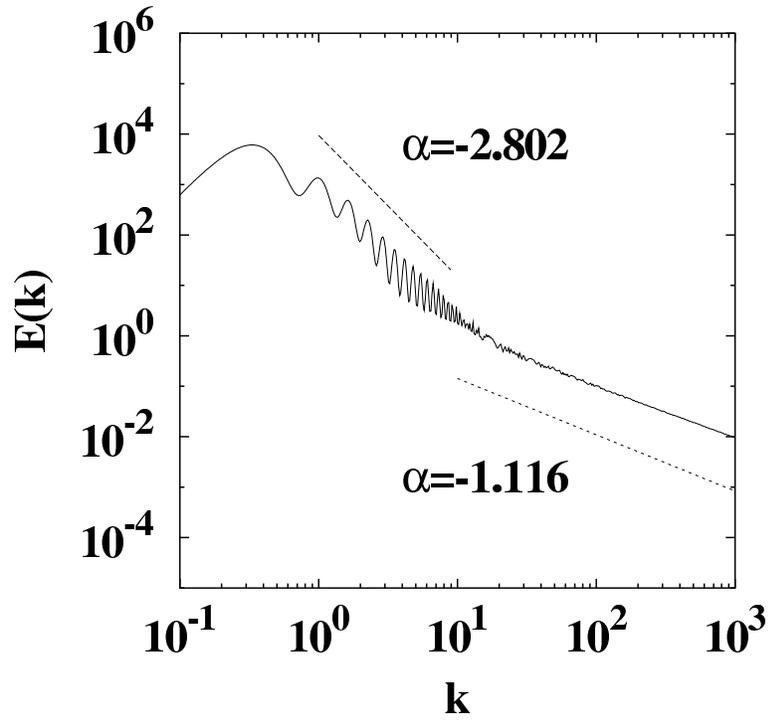}
\caption{
The power law of the energy spectrum after merging. $E(k)\sim k^{-2.8}$
 in the range $10^0<k<10^1$. The power $-1$ range is originated
 in the self energy.
}
\label{Fig:Beki}
\end{center}
\end{figure}

The energy spectrum shows the power law $E(k) \sim k^{-2.8}$ in the range 
 $10^{0}<k<10^1$, and $E(k) \sim k^{-1.1}$
 in the range $k>10^1$ (Fig.\ref{Fig:Beki}). 
The origin of the power $-1$ region is the self energy
 (the first term of eq.(\ref{eq:ES})). 
The former power is near to the value of the prediction for
 the 2D turbulence,~\cite{Ranryuu0,Ranryuu1,Ranryuu2,Ranryuu3}
 i.e., $E(k)\sim k^{-3}$.
However, the other initial condition leads to
 different values, namely $E(k)\sim k^{-\alpha}, \alpha\approx 2.2-2.8$.
Moreover, the range of this power is changed
 for different distributions of vortices.
The range $10^0<k<10^1$ and the shape of spectrum in this range
 are an effect of the distribution of vortices in the merged clump. 

\begin{figure}[p]	
  \begin{center}
\includegraphics{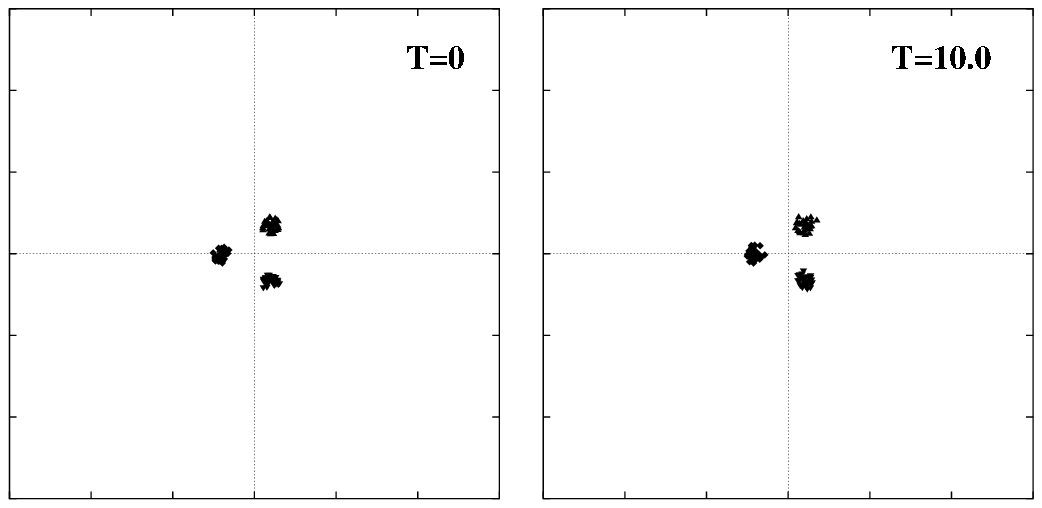}
    \caption{
The clumps without background vorticity distribution. 
Each clump contains 30 vortices.
The initial configuration of vortices of clumps are same as Fig.\ref{Fig:VrMg}.
}
\label{Fig:VrNoMg}
  \end{center}
\end{figure}

In order to confirm the effect of the background vorticity distribution,
 we calculate the system without the background
 of which configuration of vortices in the clumps
 are the same as the system with the background (Fig.\ref{Fig:VrNoMg}).
In this case, the clumps rotate preserving an equilateral triangular
 configuration for long time.
Therefore this means that the background vortices merge clumps which are 
 located separately so as not to merge without the background.

%
\section{Discussion and Conclusions}
In this paper, the vortex crystal and the vortex merger in the system of
 a few hundred point vortices are studied. The three-point vortex system
 without the background is investigated as a fundamental process
 of a three clumps system in the background vorticity distribution. 

In \S2, we studied the three-point vortex system with boundary. In order to 
 investigate chaotic behavior of this system, the Poincar\'{e} plot is 
 calculated. The size of the plotted region on the section
 is closely related to the density of states:
 in the parameter where the density of state is low,
 the plotted area on the Poincar\'{e} section is narrow and vice versa.
 The topology of the plotted area changes when the energy is varied across
 the value of the energy in which three point vortices are at the vertices of
 an equilateral triangular configuration. 
When point vortices are in an equilateral triangular configuration,
 the plots on the Poincar\'{e} section are at the origin and on
 the circle of which radius is $\sqrt{1/2}$.
When the parameters change from this value, there are no plots at the origin
 on the section.
We confirm that the sign of ``temperature'' influence the property
 of Poincar\'{e} plot. When the ``temperature'' is negative,
 the region of Poincar\'{e} plot is scooped out from inner of the region. 
The positive ``temperature'' leads to the Poincar\'{e} plot
 in which there is no Tori.
The property of the Poincar\'{e} plot reflects the motion of point vortices.
Tori or ``chaotic sea'' on this section have the relationship with
 the type of the motion of point vortices. 
The orbit that two point vortices rotate each other
 apart from the other point vortex draws tori on $(P_1,R_1)$-plane.
The cusps are seen at the locus of the motion
 in which ``chaotic sea'' emerges on the section. The former indicates
 regular, and the latter shows chaotic behavior of point vortices.

The energy spectrum of this system is introduced in \S3. Using this result,
 we calculate the energy spectrum of the $N$-point vortex system 
 in merging process in \S4.
The energy spectrum in the range around $10^0<k<10^1$
 relaxes to the form of $E(k)\sim k^{-\alpha},(\alpha\approx 2.2-2.8)$.
This range and the power are changed as the distribution of vortices is varied.
 In \S4, the vortex crystal is reproduced in the system of point vortices.
 This shows that continuous distribution of vorticity is not necessary
 for vortex crystals. 
We find the energy transfer is occurred in the process of the vortex crystal 
 of our simulation.
The energy of clumps is disgorged to the background,
 and the clumps form triangle lattice.
It is observed that the point vortices of the clumps and
 of the point vortices of the background are mixing each other.
However, the number of voritices in each clump is conserved.
This mixing effect of point vortices in the crystallization of clumps 
 must be investigated.

In our simulation, in the process of the vortex crystal,
 the arranging order of the clumps along the direction of rotation
 (namely, the sign of symmetry parameter $S$ introduced at \S 4.1) does not
 change; however the arranging order of the point vortices are changed in 
 a three-point vortex system as Fig.\ref{Fig:PS-VC}(O-1).
When this order of the clumps is permutated,
 vortex crystals does not occur in our simulations.
This suggests a possibility that there are a relation between
 the crystallization of three clumps
 and the motion of three point vortices mentioned above. 
In a three-point vortex system, three vortices change the arranging order
 at some initial conditions and do not change at the other.
We suppose that there are at least three factors for the crystallization
 of clumps, namely the parameters $(I,E)$,
 the ratio of circulation $\Gamma_{\text{background}}/\Gamma_{\text{total}}$, 
 and the ratio of the number of vortices
 $N_{\text{background}}/N_{\text{total}}$.
Of course, the location of clumps also influences the crystallization. 
It must be investigated which parameter restrain the change of the order
 of rotating vortices. 
%
\section*{Acknowledgments}
We are indebted to Prof. H. Tomita, Prof. H. Hayakawa and Dr. Y. Yatsuyanagi
 for valuable discussions about this work. We also acknowledge
 Prof. Y. Kiwamoto and Dr. A. Sanpei for inspiration of this work. 
\section*{Appendix A}
\begin{figure}[p]
\begin{center}
\includegraphics[width=10.0cm]{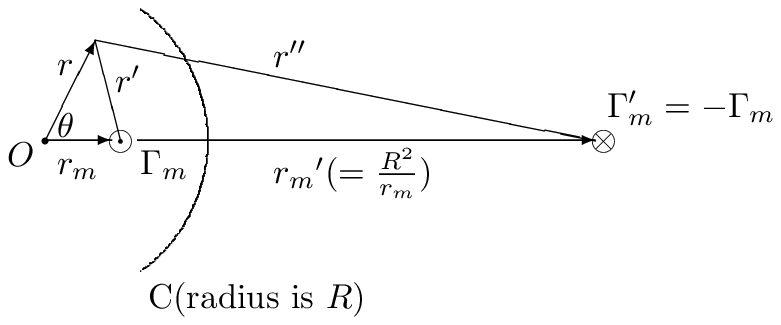}
\end{center}
\caption{The image vortex}
\label{fig:ImVor}
\end{figure}

\setcounter{equation}{0}
\setcounter{section}{1}
\renewcommand{\thesection}{\Alph{section}}
\renewcommand{\theequation}{\thesection$\cdot$\arabic{equation}}
We want to calculate $v_{\theta}=-\partial \Psi/\partial r$
 which is the second term of the vorticity field in eq.(\ref{eq:IdcVor});
the term of induced vortex sheet on the circular boundary.
When the $m$-th vortex is at the position $\bm{r}_m=(r_m,0)$,
using the method of images, the stream function of this vortex at the position
 $\bm{r}=(r,\theta)$ is 
\begin{align}
\Psi_m(r,\theta)=&-\frac{\Gamma_m}{2\pi}\log r^{\prime} 
+\frac{\Gamma_m}{2\pi}\log r^{\prime\prime}
-\frac{\Gamma_m}{2\pi}\log{\frac{R}{r_m}} \notag \\
=&-\frac{\Gamma_m}{4\pi}\log(r^2+{r_m}^2-2rr_m \cos\theta) 
+\frac{\Gamma_m}{4\pi}\log(r^2+{{r_m}^{\prime}}^2
-2r{r_m}^{\prime}\cos\theta)-\frac{\Gamma_m}{2\pi}\log{\frac{R}{r_m}} \notag \\
=&-\frac{\Gamma_m}{4\pi}\log(r^2+{r_m}^2-2rr_m \cos\theta) \notag \\
&+\frac{\Gamma_m}{4\pi}
\log\left[r^2+\left(\frac{R^2}{r_m}\right)^2
-2r\left(\frac{R^2}{r_m}\right)\cos\theta\right]-\frac{\Gamma_m}{2\pi}\log{\frac{R}{r_m}},
\end{align}
where in the last line, the relation ${r_m}^{\prime}=R^2/r_m$ is used.
$r^{\prime}$, $r^{\prime\prime}$ and ${r_m}^{\prime}$ are,
 as given by Fig.\ref{fig:ImVor},
 the distance between $\bm{r}$ and the $m$-th vortex,
 the distance between $\bm{r}$ and the image vortex of the $m$-th vortex,
 and the distance between the origin and the image vortex,
 respectively.
 $\theta$ is the angle between $\bm{r}$ and $\bm{r}_m$.
%
Therefore $\partial \Psi_m/\partial r$ is written as
\begin{equation}
\frac{\partial\Psi_m}{\partial r}
=-\frac{\Gamma_m}{4\pi}
\left[
\frac{2r-2r_m \cos \theta}{r^2+{r_m}^2-2rr_m\cos\theta}
-\frac{2r-\frac{2R^2}{r_m}\cos\theta}
{r^2+\frac{R^4}{{r_m}^2}-2\frac{R^2}{r_m}r\cos\theta} \right].\label{eq:PR}
\end{equation}
When $r=R$, eq.(\ref{eq:PR}) becomes
\begin{align}
\left.\frac{\partial\Psi_m}{\partial r}\right|_{r=R}
&=-\frac{\Gamma_m}{4\pi}
\left[
\frac{2R-2r_m \cos \theta}{R^2+{r_m}^2-2Rr_m\cos\theta}
-\frac{2\frac{{r_m}^2}{R}-2r_m\cos\theta}{R^2+{r_m}^2-2Rr_m\cos\theta}\right] \\
&=-\frac{\Gamma_m}{4\pi}
\frac{2(R^2-{r_m}^2)}{R(R^2+{r_m}^2-2Rr_m\cos\theta)} \\
&=-\Gamma_m \frac{R^2-{r_m}^2}{2\pi R{| \bm{R}-\bm{r}_m |}^2},
\end{align}
where $\bm{R}=R\bm{r}/r$. The sum about $m$ of the above equation is 
 the second term of eq.(\ref{eq:IdcVor2}).
%
%
\section*{Appendix B}
\setcounter{equation}{0}
\setcounter{section}{2}
\renewcommand{\thesection}{\Alph{section}}
\renewcommand{\theequation}{\thesection$\cdot$\arabic{equation}}
In this Appendix, the derivation of the energy spectrum $\tilde{E}(k)$
 is shown. The starting point is 
\begin{align}
\tilde{E}(k)
&=\frac{| \tilde{\bm v}(\bm k)|^2}{2} 
=\frac{| \tilde\omega(\bm k)|^2}{2k^2} \notag \\
&=\frac{1}{2k^2}
\sum_m\sum_n\tilde{\Gamma}_m(\bm{k})\tilde{\Gamma}_n(-\bm{k})
\exp[\text{i}\bm{k}\cdot(\bm{r}_m-\bm{r}_n)],
\end{align}
where,
\begin{align}
\tilde{\Gamma}_m(\bm{k}) &= \frac{1}{2\pi}\Gamma_m\left[
1-\frac{R^2-{r_m}^2}{2\pi R}\oint_{r=R} \frac{\exp(\text{i}\bm{k}\cdot\bm{R})}
{| \bm{R}-\bm{r}_m|^2} \text{d}\bm{r} \cdot \exp(-\text{i}\bm{k}\cdot\bm{r}_m)
\right]. \label{eq:tGamma}
\end{align}
First, for the second term in eq.(\ref{eq:tGamma}),
 we use the following relation 
\begin{equation}
\begin{split}
\exp(\text{i}\bm{k}\cdot\bm{R}) &= \text{e}^{\text{i}kR\cos(\theta-\phi_m)} 
= \text{e}^{\text{i}kR\sin(\theta-\phi_m-3\pi/2)}      \\
&=\sum_{l=-\infty}^{\infty}J_l(kR)\text{e}^{\text{i}l(\theta-\phi_m-3\pi/2)},
\end{split}
\end{equation}
where $\theta$ and $\phi_m$ are the angles between $\bm{r}_m$ and $\bm{R}$,
 and between $\bm{k}$ and $\bm{r}_m$, respectively.
Therefore we have
\begin{align}
\oint_{r=R} 
\frac{\exp(\text{i}\bm{k}\cdot\bm{R})}
{| \bm{R}-\bm{r}_m|^2} \text{d}\bm{r}
&=\int_0^{2\pi} \text{d}\theta R
\frac{\sum_{l=-\infty}^{\infty}J_l(kR)\text{e}^{\text{i}l(\theta-\phi_m-3\pi/2)}}
{R^2\left[1+\left(\frac{r_m}{R}\right)^2-2\frac{r_m}{R}\cos\theta\right]} \\
&=\frac{1}{R}\sum_{l=-\infty}^{\infty}J_l(kR)(\text{i})^l
\int_0^{2\pi} \text{d}\theta
\frac{\text{e}^{\text{i}l(\theta-\phi_m)}}
{1+\left(\frac{r_m}{R}\right)^2-2\frac{r_m}{R}\cos\theta}.\label{eq.VnsSin}
\end{align}
The imaginary part of the integral in the last line of eq.(\ref{eq.VnsSin})
 vanishes because it is an odd function. The integral is evaluated as 
\begin{equation}
\int_0^{2\pi} \text{d}\theta
\frac{\text{e}^{\text{i}l\theta}}
{1+\left(\frac{r_m}{R}\right)^2-2\frac{r_m}{R}\cos\theta}
=\frac{2\pi}{1-\frac{{r_m}^2}{R^2}}\left(\frac{r_m}{R}\right)^{|l|}. 
\end{equation}
Since  ${r}_m/R<1$, the following relation can be used ($|a|<1$).
\begin{equation}
\int_0^\pi\frac{\cos\alpha x}{1-2a\cos x +a^2 }\,\text{d}x
=\frac{\pi a^{|\alpha|}}{1-a^2}.
\end{equation} 
Thus, we have
\begin{align}
\oint_{r=R}\frac{\exp(\text{i}\bm{k}\cdot\bm{R})}{| \bm{R}-{r_m}|^2}\text{d}\bm{r}
&=\frac{1}{R}\sum_{l=-\infty}^{\infty}J_l(kR)(\text{i})^l
\frac{2\pi}{1-\frac{{r_m}^2}{R^2}}\left(\frac{r_m}{R}\right)^{|l|} 
\text{e}^{-\text{i}l\phi_m}\\
&=\frac{2\pi R}{R^2-{r_m}^2}
\left[J_0(kR)
+2\sum_{l=1}^{\infty}J_l(kR)(\text{i})^l\cos(l\phi_m)
\left(\frac{r_m}{R}\right)^l  
\right].
\end{align}
Finally eq.(\ref{eq:tGamma}) is given by
\begin{equation}
\begin{split}
\tilde{\Gamma}_m(\bm{k}) 
&= \frac{1}{2\pi}\Gamma_m\left[1-\frac{R^2-{r_m}^2}{2\pi R}
\oint_{r=R} \frac{\exp(\text{i}\bm{k}\cdot\bm{R})}
{| \bm{R}-\bm{r}_m|^2} \text{d}\bm{r} \cdot \exp(-\text{i}\bm{k}\cdot\bm{r}_m)
\right]   \\
&=\frac{1}{2\pi}\Gamma_m\left[1-
\sum_{l=0}^{\infty}\varepsilon_lJ_l(kR)\cos(l\phi_m)
\left(\frac{\text{i}r_m}{R}\right)^l \exp(-\text{i}\bm{k}\cdot\bm{r}_m)
\right],
\end{split}
\end{equation}
where
\begin{equation}
\varepsilon_l= \begin{cases}
	1 & (l=0) \\
	2 & (l\geq1).
	\end{cases} 
\end{equation}
Substituting this into $|\tilde\omega(\bm k)|^2$, we get 
\begin{align}
|\tilde\omega(\bm k)|^2
&=\sum_m\sum_n\tilde{\Gamma}_m(\bm{k})\tilde{\Gamma}_n(-\bm{k})
\exp[\text{i}\bm{k}\cdot(\bm{r}_m-\bm{r}_n)] \\ 
&=\frac{1}{(2\pi)^2}\sum_m\sum_n\Gamma_m\Gamma_n
\left\{
\exp[\text{i}\bm{k}\cdot(\bm{r}_m-\bm{r}_n)]
-{\sum_l}^{(m)}-{\sum_{l^{\prime}}}^{(n)}
+{\sum_{l,l^\prime}}^{(m,n)}
\right\}.
\end{align}
Here we define
\begin{equation}
\begin{split}
{\sum_l}^{(m)}
& \equiv
\sum_{l=0}^{\infty}\varepsilon_lJ_l(kR)\cos(l\phi_m)
\left(\frac{\text{i}r_m}{R}\right)^l \exp(-\text{i}\bm{k}\cdot\bm{r}_n),   \\
{\sum_{l^{\prime}}}^{(n)}
&\equiv
\sum_{l=0}^{\infty}
 \varepsilon_{l^\prime}J_{l^\prime}(kR)\cos(l^\prime\phi_n)
 \left(\frac{-\text{i}r_n}{R}\right)^{l^\prime} \exp(\text{i}\bm{k}\cdot\bm{r}_m),    \\
{\sum_{l,l^\prime}}^{(m,n)}
& \equiv 
\sum_{l=0}^{\infty}
\sum_{l^\prime=0}^{\infty}
\varepsilon_l\varepsilon_{l^\prime} 
J_l(kR)\cos(l\phi_m) J_{l^\prime}(kR)\cos(l^\prime\phi_n)
\left(\frac{\text{i}r_m}{R}\right)^l
\left(\frac{-\text{i}r_n}{R}\right)^{l^\prime}.
\end{split}
\end{equation}
Integrating over angles $\varphi$ of $\bm{k}$,
 the energy spectrum can be written as
\begin{equation}
\begin{split}
\tilde{E}(k)&=\int_0^{2\pi}\text{d}\varphi\,k\frac{|\tilde{\bm v}(\bm k)|^2}{2}
=\frac{1}{2k}\int_0^{2\pi}\text{d}\varphi|\tilde\omega(\bm k)|^2   \\
&=\frac{1}{2(2\pi)^2k}\int_0^{2\pi}\text{d}\varphi
\sum_m\sum_n\Gamma_m\Gamma_n 
\left\{
\exp[\text{i}\bm{k}\cdot(\bm{r}_m-\bm{r}_n)]
-{\sum_l}^{(m)}-{\sum_{l^{\prime}}}^{(n)}
+{\sum_{l,l^\prime}}^{(m,n)}
\right\}\\
&\equiv (A)+(B)+(C)+(D),\label{eq:Esp}
\end{split}
\end{equation}
where,
\begin{align}
(A)=&\frac{1}{2(2\pi)^2k}\int_0^{2\pi}\text{d}\varphi
\sum_m\sum_n\Gamma_m\Gamma_n 
\exp[\text{i}\bm{k}\cdot(\bm{r}_m-\bm{r}_n)]
,\\
(B)=&-\frac{1}{2(2\pi)^2k}\int_0^{2\pi}\text{d}\varphi
\sum_m\sum_n\Gamma_m\Gamma_n 
\sum_{l=0}^{\infty}\varepsilon_lJ_l(kR)\cos(l\phi_m)
\left(\frac{\text{i}r_m}{R}\right)^l \exp(-\text{i}\bm{k}\cdot\bm{r}_n)
,\\
(C)=&-\frac{1}{2(2\pi)^2k}\int_0^{2\pi}\text{d}\varphi
\sum_m\sum_n\Gamma_m\Gamma_n 
\sum_{l=0}^{\infty}
 \varepsilon_{l^\prime}J_{l^\prime}(kR)\cos(l^\prime\phi_n)
 \left(\frac{-\text{i}r_n}{R}\right)^{l^\prime} \exp(\text{i}\bm{k}\cdot\bm{r}_m)
,\\
(D)=&\frac{1}{2(2\pi)^2k}\int_0^{2\pi}\text{d}\varphi
\sum_m\sum_n\Gamma_m\Gamma_n  \notag \\
&\times
\left\{
\sum_{l=0}^{\infty}
\sum_{l^\prime=0}^{\infty}
\varepsilon_l\varepsilon_{l^\prime} 
J_l(kR)\cos(l\phi_m) J_{l^\prime}(kR)\cos(l^\prime\phi_n)
\left(\frac{\text{i}r_m}{R}\right)^l\left(\frac{-\text{i}r_n}{R}\right)^{l^\prime}
\right\}.
\end{align}
We calculate each term as follows.

In order to calculate the first term $(A)$,
suppose that the angle between $\bm{k}$ and $(\bm{r}_m-\bm{r}_n)$ is $\psi$.
 We can expand $\exp[\text{i}\bm{k}\cdot(\bm{r}_m-\bm{r}_n)]$
 in the series of the Bessel functions,
\begin{equation}
\exp[\text{i}\bm{k}\cdot(\bm{r}_m-\bm{r}_n)]=\exp[\text{i}k|\bm{r}_m-\bm{r}_n|\cos\psi]
=\sum_{l=-\infty}^{\infty}J_l(k|\bm{r}_m-\bm{r}_n|)\text{e}^{\text{i}l(\psi+\pi/2)}. 
\end{equation}
Thus we have
\begin{align}
\int_0^{2\pi}\text{d}\varphi
\exp[\text{i}\bm{k}\cdot(\bm{r}_m-\bm{r}_n)]
&=\int_0^{2\pi}\text{d}\varphi  
\sum_{l=-\infty}^{\infty}J_l(k| \bm{r}_m-\bm{r}_n|)\text{e}^{\text{i}l(\psi+\pi/2)}\notag \\
&=\sum_{l=-\infty}^{\infty}J_l(k| \bm{r}_m-\bm{r}_n|)\int_0^{2\pi}\text{d}\varphi
\,\text{e}^{\text{i}l(\psi+\pi/2)} \notag\\
&=2\pi J_0(k| \bm{r}_m-\bm{r}_n|),
\end{align}
where $\psi=\chi-\varphi$ and $\chi$ is the angle of $\bm{r}_m-\bm{r}_n$.
Then, the first term $(A)$ is written as
\begin{equation}
\begin{split}
(A) &= \frac{1}{4\pi k}
\sum_m\sum_n\Gamma_m\Gamma_nJ_0(k| \bm{r}_m-\bm{r}_n|)\\
&=\frac{1}{4\pi k}\left[\sum_n{\Gamma_n}^2
+2\sum_m\sum_{m<n}\Gamma_m\Gamma_nJ_0(k| \bm{r}_m-\bm{r}_n|) \right].
\end{split}
\end{equation}

The second term $(B)$ of eq.(\ref{eq:Esp}) is 
\begin{equation}
(B)
=-\frac{1}{2(2\pi)^2 k}\int_0^{2\pi}\text{d}\varphi\sum_m\sum_n\Gamma_m\Gamma_n
\sum_{l=0}^{\infty}\varepsilon_lJ_l(kR)\cos(l\phi_m)
\left(\frac{\text{i}r_m}{R}\right)^l \exp(-\text{i}\bm{k}\cdot\bm{r}_n) \label{eq:2term}.
\end{equation}
Suppose that $\varphi$ and $\varphi_n$ are the angle components
 of $\bm{k}$ and $\bm{r}_n$ respectively. 
We have 
\begin{equation}
 \exp(-\text{i}\bm{k}\cdot\bm{r}_n)
 =\sum_{l=-\infty}^{\infty}J_l(kr_n)
\text{e}^{-\text{i}l\left(\varphi-\varphi_n+\frac{\pi}{2}\right)}
\end{equation}
and
\begin{equation}
\cos(l\phi_n)
=\frac{1}{2}\left\{\text{e}^{\text{i}l(\varphi-\varphi_n)}+\text{e}^{-\text{i}l(\varphi-\varphi_n)}
 \right\},
\end{equation}
where $\phi_n=\varphi-\varphi_n$.
Substituting these into the sum over $l$ in eq.(\ref{eq:2term}), we obtain
\begin{align}
&\sum_{l=0}^{\infty}  \varepsilon_lJ_l(kR)\cos(l\phi_m)
\left(\frac{\text{i}r_m}{R}\right)^l \exp(-\text{i}\bm{k}\cdot\bm{r}_n)    \notag \\
&=\frac{1}{2}
\sum_{l=0}^{\infty}\varepsilon_l\left(\frac{\text{i}r_m}{R}\right)^lJ_l(kR)
\sum_{l^\prime=-\infty}^{\infty}J_{l^\prime}(kr_n)
\text{e}^{-\text{i}l^\prime\left(\varphi-\varphi_n+\frac{\pi}{2}\right)}
\left\{\text{e}^{\text{i}l(\varphi-\varphi_m)}+\text{e}^{-\text{i}l(\varphi-\varphi_m)} \right\}.
\label{eq:lpart}
\end{align}
Integrating over the angle $\varphi$,
 the sum over $l^{\prime}$ in eq.(\ref{eq:lpart}) is written as
\begin{align}
&\sum_{l^\prime=-\infty}^{\infty}
\int_0^{2\pi}\text{d}\varphi \, J_{l^\prime}(kr_n)
\text{e}^{-\text{i}l^\prime\left(\varphi-\varphi_n+\frac{\pi}{2}\right)}
\left\{\text{e}^{\text{i}l(\varphi-\varphi_m)}+\text{e}^{-\text{i}l(\varphi-\varphi_m)} \right\}  \notag\\
&=J_l(kr_n)
(-\text{i})^l\cdot 2\cos[l(\varphi_m-\varphi_n)] \cdot 2\pi,
\end{align}
where the integral over $\varphi$ of $\text{e}^{\text{i}\varphi(l-l^{\prime})}$
 is zero when $l\neq l^{\prime}$ and is $2\pi$ when $l=l^{\prime}$.
Then we can write the second term $(B)$ as
\begin{align}
(B)
&=-\frac{1}{4\pi k}\sum_m\sum_n\Gamma_m\Gamma_n
\sum_{l=0}^{\infty}\varepsilon_l\left(\frac{r_m}{R}\right)^lJ_l(kR)J_l(kr_n)
\cdot \cos[l(\varphi_m-\varphi_n)] .
\end{align}

For the third term $(C)$,
 a similar calculation to the second term $(B)$ can be applied.
Then we have
\begin{equation}
(C)=(B) 
=-\frac{1}{4\pi k}\sum_m\sum_n\Gamma_m\Gamma_n
\sum_{l=0}^{\infty}\varepsilon_l\left(\frac{r_m}{R}\right)^lJ_l(kR)J_l(kr_n)
\cdot \cos[l(\varphi_m-\varphi_n)] 
\end{equation}

The last term $(D)$ is 
\begin{align}
(D)&=\frac{1}{2(2\pi)^2 k}\sum_m\sum_n\Gamma_m\Gamma_n  \notag \\
&\times\sum_{l=0}^{\infty}\sum_{l^\prime=0}^{\infty} \int_0^{2\pi}\text{d}\varphi
\varepsilon_l\varepsilon_{l^\prime}
J_l(kR)J_{l^\prime}(kR)\cos(l\phi_m)\cos(l^\prime\phi_n)
\left(\frac{\text{i}r_m}{R}\right)^l\left(\frac{-\text{i}r_n}{R}\right)^{l^\prime}.
\end{align}
Since $l$ and $l^\prime$ are both integers ( $l,l^\prime \geq 0$ ),
 then we have
\begin{align}
&\int_0^{2\pi}\text{d}\varphi
\cos(l\phi_m)\cos(l^\prime\phi_n)
=\int_0^{2\pi}\text{d}\varphi
\cos[l(\varphi-\varphi_m)]\cos[l^\prime(\varphi-\varphi_n)]  \notag  \\
&= \begin{cases}
2\pi & (l=l^\prime=0)  \\
\pi\cos[l(\varphi_m-\varphi_n)]  & (l=l^\prime\neq 0)\\
0 & (\text{otherwise}).
\end{cases}
\end{align}
Thus the fourth term $(D)$ is given by
\begin{align}
(D)&=\frac{1}{4\pi k}\sum_m\sum_n\Gamma_m\Gamma_n
\sum_{l=0}^{\infty} \varepsilon_l
\left[J_l(kR)\right]^2
\left(\frac{r_mr_n}{R^2}\right)^l\cos[l(\varphi_m-\varphi_n)] .
\end{align}
Substituting these four terms into eq.(\ref{eq:Esp}),
 the formula for the energy spectrum is given by
\begin{equation}
\begin{split}
\tilde{E}(k)=&\frac{1}{4\pi k} \left\{ \sum_n{\Gamma_n}^2  \right.  
+2\sum_m\sum_{n<m}\Gamma_m\Gamma_nJ_0(k| \bm{r}_m-\bm{r}_n|) \\
&-2\sum_m\sum_n\Gamma_m\Gamma_n
\sum_{l=0}^{\infty}\varepsilon_l\left(\frac{r_m}{R}\right)^lJ_l(kR)J_l(kr_n)
\cdot \cos[l(\varphi_m-\varphi_n)] \\
&+\sum_m\sum_n\Gamma_m\Gamma_n
\sum_{l=0}^{\infty} \varepsilon_l
\left[J_l(kR)\right]^2
\left . \left(\frac{r_mr_n}{R^2}\right)^l\cos[l(\varphi_m-\varphi_n)]   \right\}.
\end{split}
\end{equation}

\end{document}